\documentclass[twocolumn]{aastex701}

\begin{document}

\title{The 200--300 GHz Survey for 18 Class II Disks in the Ophiuchus Star-Forming Complex}

\author[0009-0007-3677-8040]{Chia-Ying Chung}
\affiliation{Department of Physics, National Sun Yat-Sen University, No. 70, Lien-Hai Road, Kaohsiung City 80424, Taiwan, R.O.C.}
\affiliation{Department of Physics and Astronomy, Johns Hopkins University, 3400 N Charles Street, Baltimore, MD 21218, USA}  
\email{sharon0311chung@gmail.com}

\author[orcid=0000-0003-2300-2626]{Hauyu Baobab Liu}
\affiliation{Department of Physics, National Sun Yat-Sen University, No. 70, Lien-Hai Road, Kaohsiung City 80424, Taiwan, R.O.C.}
\affiliation{Center of Astronomy and Gravitation, National Taiwan Normal University, Taipei 116, Taiwan}  
\email[show]{hyliu.nsysu@g-mail.nsysu.edu.tw}
\correspondingauthor{Hauyu Baobab Liu}

\author[0009-0004-7545-1735]{Chuan-Yung Lin}
\affiliation{Department of Physics, National Sun Yat-Sen University, No. 70, Lien-Hai Road, Kaohsiung City 80424, Taiwan, R.O.C.}
\email{arstin001@gmail.com}

\author[orcid=0009-0003-4821-5502]{Linjing Feng}
\affiliation{National Astronomical Observatories, Chinese Academy of Sciences, 20A Datun Road, Chaoyang District, Beijing 100012, China}
\affiliation{University of Chinese Academy of Sciences, Beijing 100049, China}
\email{ljfeng@nao.cas.cn}

\author[0000-0003-2253-2270]{Sean M. Andrews}
\affiliation{Center for Astrophysics \textbar\, Harvard \& Smithsonian, 60 Garden St., Cambridge, MA 02138, USA}
\email{sandrews@cfa.harvard.edu}

\author[0000-0003-0685-3621]{Mark A. Gurwell}
\affiliation{Center for Astrophysics \textbar\, Harvard \& Smithsonian, 60 Garden St., Cambridge, MA 02138, USA}
\email{mgurwell@cfa.harvard.edu}

\author[0000-0002-9154-2440]{Melvyn Wright}
\affiliation{Department of Astronomy, University of California, Berkeley, 501 Campbell Hall, Berkeley, CA 94720-3441, USA}
\email{wright@astro.berkeley.edu}


\begin{abstract}
We measured the 198--310 GHz flux densities of 18 Class II protoplanetary disks in the Ophiuchus star-forming complex using the Submillimeter Array (SMA).
The $\sim$200--300 GHz spectral indices ($\alpha_{200-300}$) of these disks are 2.3$^{+0.2}_{-0.5}$.
In this observed frequency range, emission properties of the Ophiuchus disks are very similar to the 48 spatially compact Class II disks in the Taurus-Auriga region observed in a previous SMA survey.
Combining the Ophiuchus and Taurus-Auriga samples, the Pearson correlation coefficient between $\alpha_{200-300}$ and the flux density at 250 GHz ($F_{\rm 250 GHz}$) is 0.27 ($p=0.02$).
We do not find a convincing correlation between  $\alpha_{200-300}$ and the $R_{\rm 95\%}$ radius of the disks. 
Our interpretation for the $\alpha_{200-300}$ values being close to 2 in the spatially compact disks is that at $\gtrsim$200 GHz frequencies, the continuum emission in these disks is dominated by optically thick dust thermal emission, independent of $F_{\rm 250 GHz}$ and $R_{\rm 95\%}$. As a result, the (sub)millimeter luminosity and the apparent disk size are positively correlated, which has been found in numerous resolved emission surveys of Class II disks. In young clusters with ages of $\sim$1--2 Myrs, Class II disks with smaller radii or lower (sub)millimeter luminosities are not necessarily those in which the dust mass budget has been consumed or dispersed. 
\end{abstract}


\section{Introduction} \label{sec:intro}
Some recent observational studies have indicated the $\sim$100 $\mu$m maximum dust grain sizes ($a_{\rm max}$) in protoplanetary disks (e.g., \citealt{Kataoka2016ApJ...820...54K,Stephens2017ApJ...851...55S,Liu2019ApJ...877L..22L,Guidi2022A&A...664A.137G,Hu2025ApJ...981...12H,Liu2026A&A...708A..95L}).
Such modest $a_{\rm max}$ values may indicate that the dust particles in the observed regions are fragile (e.g. \citealt{,Jiang2024A&A...682A..32J}) or are limited by the bouncing and/or inward migration barrier(s) of dust growth, and thus are hard to coagulate to larger sizes (for a review see \citealt{Birnstiel2023arXiv231213287B}). 
If this is indeed the case, disks may tend to preserve their initial dust mass budget instead of consuming most of it in planet formation. 
Constraining the overall dust mass in individual protoplanetary disks can serve as a complementary test of this idea. 
Such constraints can be obtained by performing multi-wavelength observations in the (sub)millimeter region (for a pedagogical review and an overview for recent studies, see \citealt{Hildebrand1983QJRAS..24..267H,Andrews2020ARA&A..58..483A}).

\begin{deluxetable*}{lccccccccc}
\tabletypesize{\footnotesize}
\tablecolumns{10}
\tablewidth{0pt}
\tablecaption{Properties of the host protostars and the (sub)millimeter disks \label{tab:protostars}}
\tablehead{ 
\colhead{Source} & \colhead{2MASS Name} & \colhead{RA} & \colhead{Dec} & \colhead{Distance} & \colhead{Spectral Type} & \colhead{Multiplicity} & \colhead{Separation} & \colhead{$R_{\rm 95\%, disk}$}  & \colhead{Group} \\
\colhead{} & \colhead{} & \colhead{(hh:mm:ss)} & \colhead{(dd$^\circ$mm$'$ss$''$)} & \colhead{(pc)} & \colhead{} & \colhead{} & \colhead{($''$)} & \colhead{($''$)} & \colhead{} \\
\colhead{(1)} & \colhead{(2)} & \colhead{(3)} & \colhead{(4)} & \colhead{(5)} & \colhead{(6)} & \colhead{(7)} & \colhead{(8)} & \colhead{(9)} & \colhead{(10)}
}
\startdata
DoAr 16 (Haro~1-4)   & J16251052-2319145 & 16:25:10.50  & $-$23$^\circ$19$'$15.$''$1  & 135 & K4   & binary & 0.72 & 0.15 & 1 \\
AS 206 (SR~4)   & J16255615-2420481 & 16:25:56.16 & $-$24$^\circ$20$'$48.$''$7  & 135 & K4.5 &  single  & \nodata & 0.23 & 2 \\
GSS 26    & J16261033-2420548 & 16:26:10.33 & $-$24$^\circ$20$'$54.$''$8   & 139 & M0   &  single  & \nodata & 0.46 & 2 \\
VSSG 1 (Elias~2-20)   & J16261886-2428196 & 16:26:18.87 & $-$24$^\circ$28$'$20.$''$1  & 138 & M0   &  single  & \nodata & 0.38 & 2 \\
DoAr 24E  & J16262335-2420597 & 16:26:23.36 & $-$24$^\circ$21$'$0.$''$0    & 126 & G6   &  binary  & 2.06  & Ea: 0.098  & 1 \\
          &                   &             &                              &     &      &          &       & Eb: 0.024  & 1 \\
DoAr 25   & J16262367-2443138 & 16:26:23.68 & $-$24$^\circ$43$'$14.$''$3  & 138 & K5   &  single  & \nodata & 1.1 & 2 \\
WSB 31 (Elias~2-24)   & J16262407-2416134 & 16:26:24.08 & $-$24$^\circ$16$'$13.$''$8  & 136 & K6   &  single  & \nodata & 0.98 & 2 \\
GSS 39 (Elias~2-27)   & J16264502-2423077 & 16:26:45.02 & $-$24$^\circ$23$'$8.$''$3   & 139 & M0   &  single  & \nodata & 1.28 & 1 \\
YLW 8 (SR~21 or Elias~2-30)     & J16271027-2419127 & 16:27:10.27 & $-$24$^\circ$19$'$13.$''$1  & 136 & G1   & binary & 6.4 & 0.59 & 1 \\
IRS 37    & J16271757-2428562 & 16:27:17.58 & $-$24$^\circ$28$'$56.$''$2   & 139 & M4   &  single  & 1.92  & 0.099 & 2 \\
IRS 39    & J16271848-2429059 & 16:27:18.38 & $-$24$^\circ$29$'$4.$''$0   & 139 & M1.5 &  single  & \nodata & 0.090 & 3 \\
IRS 41    & J16271921-2428438 & 16:27:19.18 & $-$24$^\circ$28$'$40.$''$9  & 139 & K7   &  single  & \nodata & 0.13 & 2 \\
YLW 47    & J16273832-2436585 & 16:27:38.31 & $-$24$^\circ$36$'$59.$''$0   & 140 & K5.5 &  single  & \nodata & 0.13 & 3 \\
DoAr 33   & J16273901-2358187 & 16:27:39.00  & $-$23$^\circ$58$'$19.$''$1  & 140 & K6   &  single  & \nodata & 0.19 & 1 \\
IRS 51    & J16273982-2443150 & 16:27:39.51 & $-$24$^\circ$43$'$14.$''$6  & 139 & K6   &  single  & 1.65 & 0.17 & 2 \\
WSB 60    & J16281650-2436579 & 16:28:16.50  & $-$24$^\circ$36$'$58.$''$5  & 137 & M4   &  single  & \nodata & 0.51 & 3 \\
HBC 266   & J16284527-2428190 & 16:28:45.26 & $-$24$\circ$28$'$19.$''$4   & 82  & M3.75&  single  & 0.43  & 0.39 & 3 \\
DoAr 44   & J16313346-2427372 & 16:31:33.46 & $-$24$^\circ$27$'$37.$''$6  & 146 & K3   &  single  & \nodata & 0.47 & 3 \\
\enddata
\tablecomments{
The source name in Column 1 is adopted from the name in \citet{Andrews2007ApJ...671.1800A}. 
The 2MASS name and distance of the source in Columns 2 and 5 are from \citet{Ciezza2019MNRAS.482..698C} and the Right Ascension and Declination in columns 3 and 4 are from Gaia DR3 (\citealt{Gaia2016,Gaia2023}). 
The spectral type and multiplicity in Columns 6 and 7 are from \citet{Ratzka2005} and \citet{Ciezza2019MNRAS.482..698C}.  
The separation in column 8 are mostly quoted from \citet{Dasgupta2025ApJ...981L...4D}.
The disk size in column 9 are detailed in Appendix \ref{appdx:radii}.
} 
\end{deluxetable*}

The recent Submillimeter Array (SMA; \citealt{Ho2004ApJ...616L...1H}) surveys on 57 Class II disks in the Taurus-Auriga region ( \citealt{Chung2024ApJS..273...29C,Chung2026arXiv260805445C}) have shown that, except for a handful of exceptionally spatially extended disks, the 200--400 GHz (0.75--1.5 mm) spectral indices are concentrated to an extremely narrow region of 2.0$\pm$0.2. 
Based on their population synthesis, the simplest interpretation of the observed spectral indices is that dust thermal emission in the selected Class II samples is optically thick in this frequency range, with optical depths $\gtrsim$5. 
The high dust optical depths in these Class II disks also naturally explain the well-known relation between millimeter luminosity and disk size (\citealt{Tripathi2017ApJ...845...44T,Andrews2018ApJ...865..157A,Hendler2020ApJ...895..126H}).
Assuming the standard DSHARP dust opacities (\citealt{Birnstiel2018ApJ...869L..45B}), based on the dust optical depths inferred from the multi-wavelength observations and the disk sizes measured in higher angular resolution images, \citet{Chung2024ApJS..273...29C} found that the lower limits of dust masses in their observed Class II disks may be, in general, greater than 100 $M_{\oplus}$.
This corresponds to the initial dust mass budget of a protoplanetary disk around a $\sim$0.3 $M_{\odot}$ protostar, if we assume an initial stellar-to-disk mass ratio of 10 (i.e., marginally gravitationally unstable; \citealt{Kratter2016}) and a gas-to-dust mass ratio of 100 (\citealt{Draine2011piim.book.....D}).

\begin{deluxetable*}{cccccccccc}
\tabletypesize{\scriptsize}
\tablecolumns{9}
\setlength{\tabcolsep}{1pt}
\tablewidth{0pt}
\tablecaption{SMA observations \label{tab:obs_summary}}
\tablehead{ 
\colhead{Track ID} & \colhead{UTC Date} & \colhead{LSB Freq.} & \colhead{USB Freq.} & \colhead{Array Config.} & \colhead{uv--range} & \colhead{Flux Calib.} & \colhead{Passband Calib.} & \colhead{Gain Calib.} & \colhead{$\tau_{\rm 225 GHz}$}\tablenotemark{a} \\
\colhead{} & \colhead{(YYYY-MM-DD)} & \colhead{(GHz)} & \colhead{(GHz)} & \colhead{} & \colhead{(k$\lambda$)} & \colhead{} & \colhead{} & \colhead{} & \colhead{} 
}
\startdata 
\multicolumn{9} {c} {Project: 2021B-A011, PI: Chia-Ying Chung} \\
270 GHz–2a & 2022-05-23 & 254--266 & 274--286 & EXT & 30--200 & Titan & 3C279 & J1626-298 & 0.05--0.1 \\
          &            & 286--298 & 306--318 &     &      &        &      & \\
230 GHz–3b & 2022-06-06 & 193--205 & 213--225 & EXT & 20--160 & Titan & 3C279 & J1626-298 & $\sim$0.1 \\
          &            & 225--234 & 242--254 &     &      &        &      & \\
\hline\multicolumn{9} {c} {Project: 2023A-A009, PI: Chia-Ying Chung} \\
270 GHz-1 & 2023-08-29 & 254--266 & 274--286 & COM & 5--75 & Titan & BL~Lac & J1733-130 & $\sim$0.05 \\
          &            & 286--298 & 306--318 &     &      &        &      & \\
230 GHz-2 & 2023-08-28 & 193--205 & 213--225 & COM & 5--50 & Titan & BL~Lac & J1733-130 & $\sim$0.04 \\
          &            & 225--234 & 242--254 &     &      &        &      & \\
\enddata
\tablenotetext{a}{The atmospheric optical depth measured at 225 GHz.}
\end{deluxetable*}

Whether or not Class II disks in other molecular clouds tend to preserve their initial dust mass budgets remains uncertain. 
To address this question, we extended the SMA survey to Class II disks in the Ophiuchus molecular cloud.
The ages of the Class II disks in this molecular cloud are $\sim$1--2 Myrs, which are similar to those in the Taurus-Auriga region (see Figure 5 of \citealt{Manara2023ASPC..534..539M}).
Therefore, the comparison between the Ophiuchus and Taurus-Auriga disks manifests the cloud-to-cloud difference (e.g., due to different UV radiation fields or cosmic ionization rates) rather than evolutionary effects. 

The details of our observations and target source selections are introduced in Section \ref{sec:observation}.
Our procedures for data calibration, imaging, measuring flux densities, and deriving spectral indices are provided in Section \ref{sec:reduction}.
The results are given in Section \ref{sec:results}.
Section \ref{sec:discussion} summarizes our results with a brief discussion of the physical implications.
Appendix \ref{appdx:image} provides the images produced from our SMA observations at 286--294 GHz.
Appendix \ref{appdx:flux} provides our flux density measurements.
Appendix \ref{appdx:radii} describes how we estimated the radius for one of the target sources, DoAr~24Eb, and reveals from which references we quoted the disk radii of the other sources.

\section{Observations}\label{sec:observation}
\subsection{Sample selection}\label{sub:source}
We selected the sample from the 48 young stellar objects (YSOs) observed by \citet{Andrews2007ApJ...671.1800A}.
We only included objects with flux densities greater than 45 mJy at $\sim$350 GHz frequency ($\sim$850 $\mu$m wavelength).
To save slewing time, we further excluded the four sources, RNO~90, Wa~Oph~6, AS~209 and AS~205, which are separated from the geometry center of the other sample by $>$2$^{\circ}$.
This yields a sample of 18 Class II disks.
With an integration time of one hour, the selected objects can be detected at $>$10-$\sigma$ significance in every 12 GHz intermediate frequency (IF) of the selected SMA correlator settings, thereby meeting our target sensitivity (more details are provided in Section \ref{sub:obs}).
Table \ref{tab:protostars} summarizes the properties of the host protostars and the disk radii derived in the high angular resolution observations of dust continuum emission.

\subsection{The SMA observations}\label{sub:obs}
We conducted an SMA survey of the selected sample (Table \ref{tab:protostars}) between 2022 and 2023 (Table \ref{tab:obs_summary}). 
The observations in 2022 were taken in the extended array configuration; the follow-up observations in 2023 were taken in the compact configuration. 
Concatenating the extended and compact array configuration observations yielded 0\farcs8--1\farcs2 synthesized beam full width at half maximum (FWHM) and $\sim$40$''$ maximum recoverable angular scales. 

The observational logistics, including integration time allocation, target grouping and slewing path optimization, and the interleaving of target-calibrator scans, were designed following the strategy of \citet{Chung2024ApJS..273...29C}. 
The 18 target sources were divided into three groups based on proximity (Table \ref{tab:protostars}).
Targets in a given group were sequentially observed during the 12-minute window between complex gain calibrator scans.
This yielded a total calibration duty cycle of 15 minutes.
By looping through these three target groups every 45 minutes, we ensured that every target was visited at least 1--2 times in each track of observations. 
We observed Titan as the absolute flux calibrator and 3C279 or BL Lac as the passband calibrator at the beginning or end of each observation (Table \ref{tab:obs_summary}).

The observations were operated in dual-receiver mode, in which the local oscillator (LO) frequencies of the two receivers were tuned to 209 GHz and 238 GHz in the tracks of 230~GHz-3b and 230~GHz-2, and were tuned to 270 GHz and 302 GHz in the tracks of 270~GHz-2a and 270~GHz-1 (Table \ref{tab:obs_summary}).
The SMA Wideband Astronomical ROACH2 Machine (SWARM; \citealt{Primiani2016JAI.....541006P}) correlator simultaneously covered the $\pm$4--16 GHz intermediate frequency (IF) ranges in the upper and lower sidebands. 
The frequency coverages of the observations are summarized in Table \ref{tab:obs_summary}.

\section{Data processing}\label{sec:reduction}
\subsection{Standard SMA data calibration}\label{sub:calibration}
We perform the standard calibration of the SMA data using the MIR~IDL software package (\href{https://ui.adsabs.harvard.edu/abs/2003cdsf.conf.....C/abstract}{Qi 2003}). 
We flagged the 12--16 GHz IFs across all observations due to high noise in this range.
In the calibrations, we applied the spectral averaged system temperature ($\mathrm{T_{sys}}$) taken from wide-band measurements. 
Afterwards, we derived and applied the time-independent passband solutions.
The validity of these passband solutions was verified by inspecting the passband-calibrated spectra of the calibrators. 
For each track of observations, we calibrated the absolute fluxes for each sideband of each receiver separately. 
The complex gain phase and amplitude solutions for each sideband of each receiver were also separately derived and applied.


\subsection{Gain-phase self-calibration}\label{sub:selcal}
The self-calibration procedure includes (1) identifying target sources (hereafter the self-calibration target sources) that are bright enough for deriving self-calibration solutions in each sideband of the observations (Section \ref{sub:obs}) in short time intervals, (2) producing source models, and (3) solving and applying antenna-based, time-dependent corrections for gain phase. 
For (1) and (2), using the imaging procedure introduced in Section \ref{sub:imaging}, we first produced the preliminary images from the calibrated 254--318 GHz data (Table \ref{tab:obs_summary}) that were created in the standard calibration procedure (Section \ref{sub:calibration}).
We focused on the observations in this frequency range because they are rather sensitive to dust emission and have higher angular resolutions than the observations that were taken with the lower LO frequencies (Section \ref{sec:observation}).
From the preliminary 254--318 GHz images, we selected the following self-calibration target sources: WSB~31, GSS~39, DoAr~25, GSS~26, WSB~60, VSSG~1, YLW~8, IRS~51, DoAr~44, and AS~206.
We used the \texttt{clean} components obtained when we generated the preliminary images as source models.
Using the MIRIAD \texttt{selfcal} task (\citealt{Sault1995ASPC...77..433S}), we derived the gain-phase self-calibration solutions in 5 minutes intervals.
We visually inspected the solutions and flagged the few outliers. 
Afterwards, we employed the \texttt{scipy} \texttt{UnivariateSpline} routine (\citealt{2020SciPy-NMeth}) to interpolate the solutions in the time domain using the cubic spline function ($k=3$), and then applied the interpolated solutions to all target sources.

\subsection{Imaging}\label{sub:imaging}
We performed naturally weighted, multi-frequency synthesis (\texttt{mfs}) imaging using the MIRIAD software package (\citealt{Sault1995ASPC...77..433S}).
The observations taken in the compact and extended array configurations were jointly imaged. 
The two-dimensional continuum images were created with 600 pixels in each dimension. 
The size of a pixel is 0\farcs1. 
Our imaging procedure first created the dirty images by inversely Fourier transforming the calibrated SMA visibilities (Section \ref{sub:calibration}, \ref{sub:selcal}) using the MIRIAD \texttt{invert} task.
From the dirty images, we identified the peaks of the emission.
We then deconvolved the dirty images using the MIRIAD \texttt{clean} task for 100 iterations, restricted in the 5$''$-wide box regions around the peaks and without setting cutoff levels.
This yielded the preliminary residual images.
We measured the root-mean-square (rms) noise levels from these preliminary residual images.
To produce the final images, we deconvolved the dirty images again using the MIRIAD \texttt{clean} task, by setting the cutoff levels to 1.5 times the rms noises. 
The typically achieved synthesized beam full width at half maxima (FWHM) at the 209, 238, 270 and 302 GHz LO frequencies (Section \ref{sub:obs}) are 2\farcs1, 1\farcs7, 1\farcs2 and 1\farcs0, respectively; . 

\begin{figure*}
\hspace{-1.5cm}
\setlength{\tabcolsep}{-3pt}
    \begin{tabular}{ cccc}
    \includegraphics[width=4.9cm]{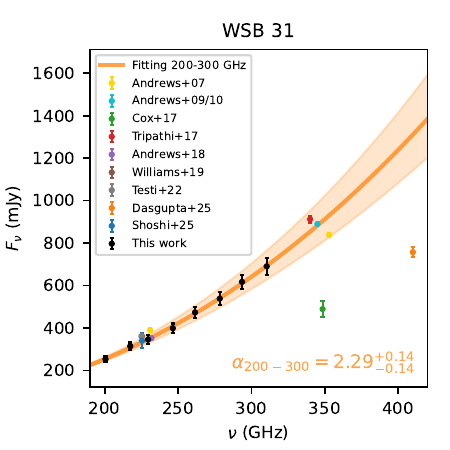} &
    \includegraphics[width=4.9cm]{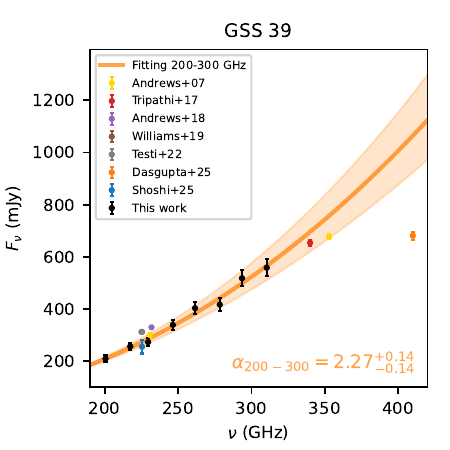} &
    \includegraphics[width=4.9cm]{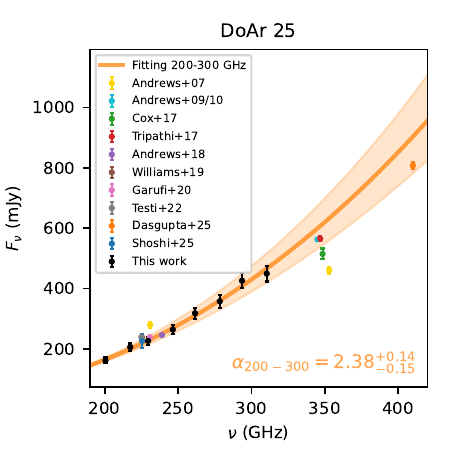} &
    \includegraphics[width=4.9cm]{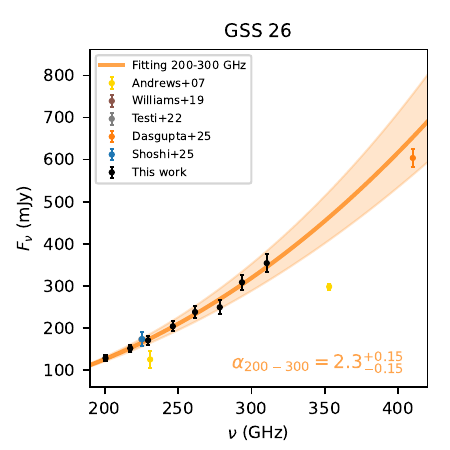} \\
    \includegraphics[width=4.9cm]{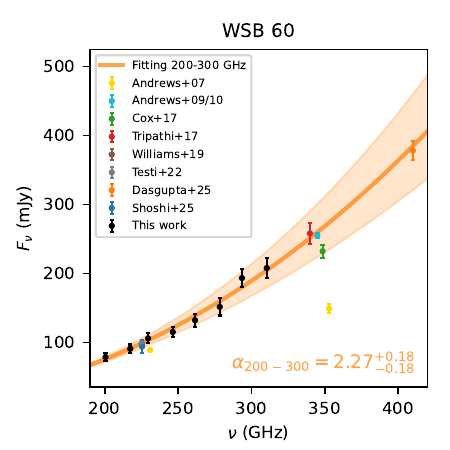} &
    \includegraphics[width=4.9cm]{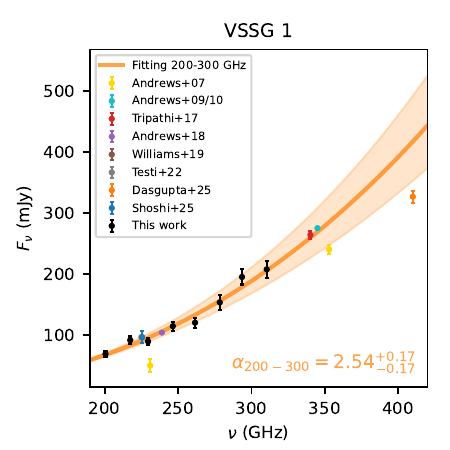} &
    \includegraphics[width=4.9cm]{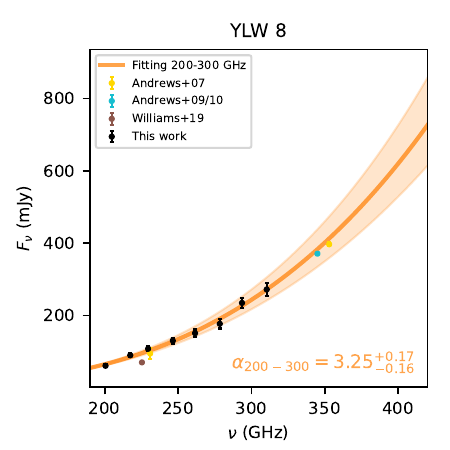}  &
    \includegraphics[width=4.9cm]{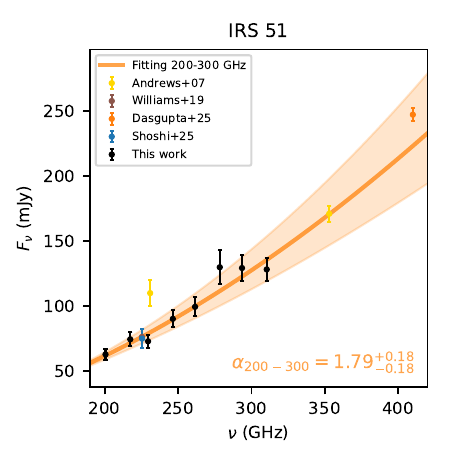} \\
    \includegraphics[width=4.9cm]{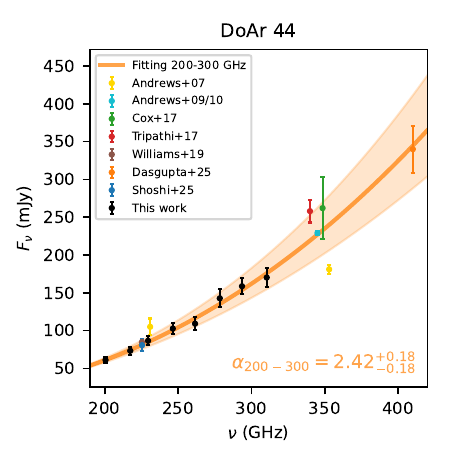} &
    \includegraphics[width=4.9cm]{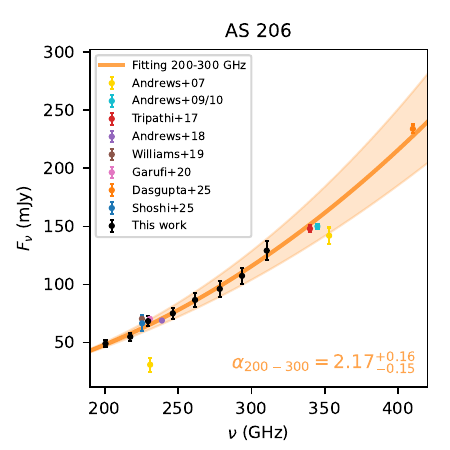} &
    \includegraphics[width=4.9cm]{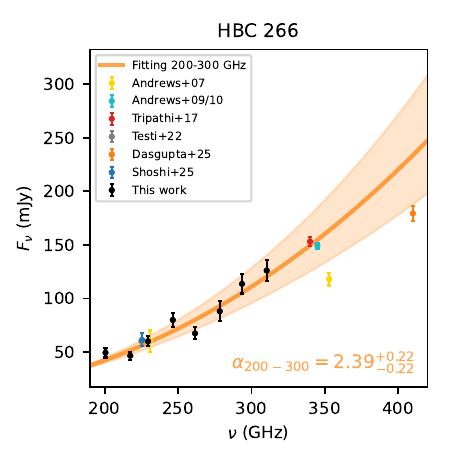} &
    \includegraphics[width=4.9cm]{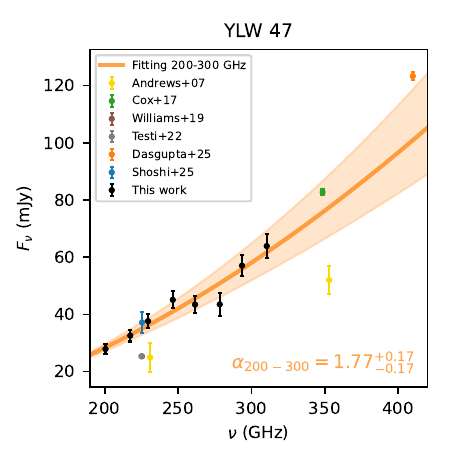} \\
    \includegraphics[width=4.9cm]{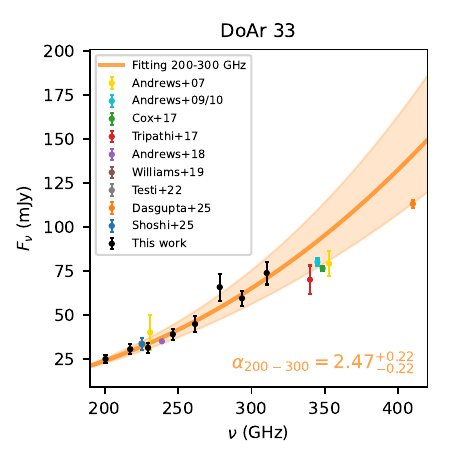} &
    \includegraphics[width=4.9cm]{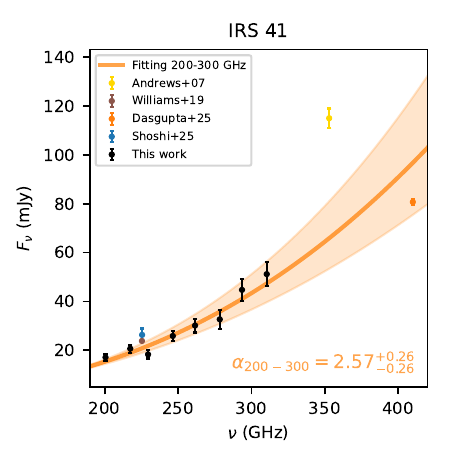} &
    \includegraphics[width=4.9cm]{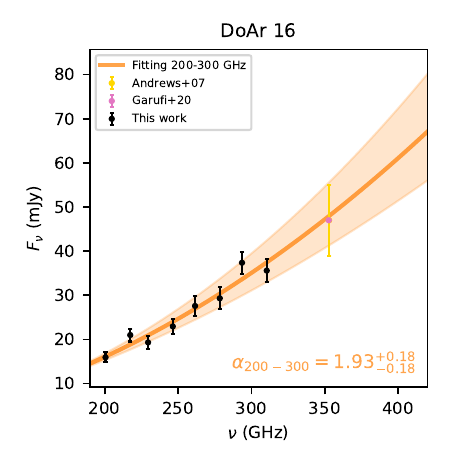} & 
    \includegraphics[width=4.9cm]{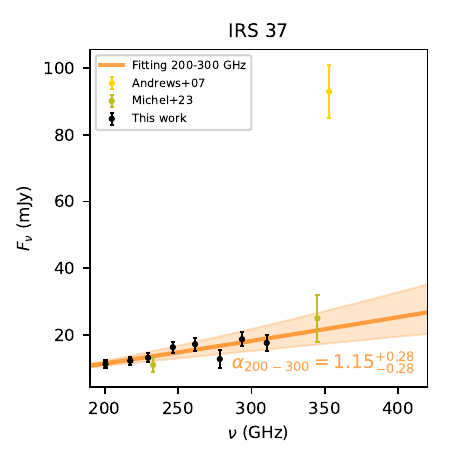} \\
    \end{tabular}
    \caption{Flux densities ($F_{\nu}$) of the selected, isolated Class II disks in the Ophiuchus region. The vertical error bars are $\pm$1$\sigma$ uncertainties ($\delta F_{\nu}$). Solid lines show the power-laws that are the best fits to the data taken from this work (Section \ref{sec:observation}); shaded areas outline the $\pm$1$\sigma$ fitting uncertainties. We compared with the observations published in \citet{Andrews2007ApJ...671.1800A, Andrews2008ApJ...678L.133A,Andrews2009ApJ...700.1502A,Andrews2010ApJ...723.1241A,Cox2017ApJ...851...83C,Tripathi2017ApJ...845...44T,Andrews2018ApJ...869L..41A,Williams2019ApJ...875L...9W, Garufi2020A&A...633A..82G,Testi2022A&A...663A..98T,Michel2023AJ....166..184M,Dasgupta2025ApJ...981L...4D,Shoshi2025PASJ...77..572S} (\citet{Andrews2008ApJ...678L.133A} reported the flux density of DoAr~25, which is identical to the value that can be quoted from \citet{Andrews2009ApJ...700.1502A}. Therefore, we omit the label of this record in the figure legend.)
    }
    \label{fig:SED}
\end{figure*}

\begin{figure}
\hspace{-1.5cm}
\setlength{\tabcolsep}{-3pt}
    \begin{tabular}{ cc  }
    \includegraphics[width=4.9cm]{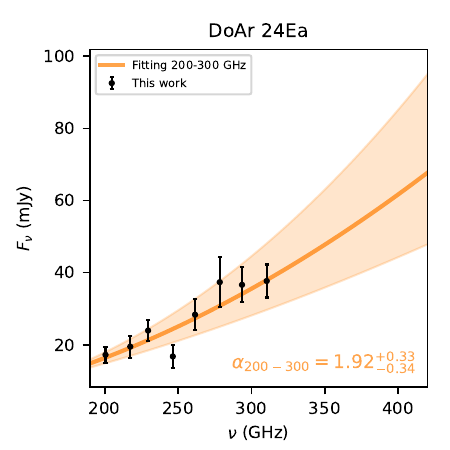} &
    \includegraphics[width=4.9cm]{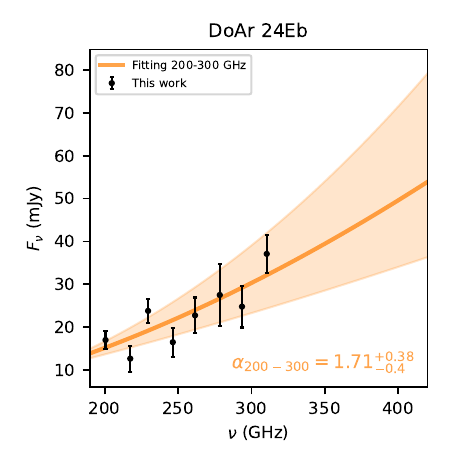} 
    \end{tabular}
    \caption{Similar to Figure \ref{fig:SED} but for the binary sources DoAr~24Ea and DoAr~24Eb.}
    \label{fig:SED_binary}
\end{figure}

\subsection{Measuring flux densities}\label{sub:flux_density}
In high angular ALMA continuum images (e.g. \citealt{Dasgupta2025ApJ...981L...4D}), all of the selected sources have $\lesssim$1$''$ radii (Table \ref{tab:protostars}).
Spatially, they were either marginally resolved or unresolved in our SMA observations with angular resolutions of 1$''$--2$''$ (Section \ref{sub:imaging}).
To measure flux densities, we performed Markov Chain Monte Carlo (MCMC) fittings in the visibility domain.
The models were chosen based on the morphology of the target sources resolved in the SMA images (more below).
The MCMC samplings were implemented using the publicly available python package \texttt{emcee} (\href{https://iopscience.iop.org/article/10.1086/670067}{Foreman-Mackey et al. 2013a}). 
The best-fit flux density ($F_{\nu}$) is taken as the value at the maximum of the posterior probability distribution, while its uncertainty ($\sigma_{F_{\nu}}$) is defined as one-half of the difference between the 16th and 84th percentiles of the posterior distribution.
We assumed a nominal absolute flux uncertainty ($\Delta$) of 5.5\%, which was realistically determined in  \citet{Chung2026arXiv260805445C} based on comparing multiple epochs of observations.
The standard error of the flux density ($\delta_{F_{\nu}}$) is $\sqrt{(\sigma_{F_{\nu}})^{2}+(F_{\nu}\times\Delta)^2}$.
The results of our flux density measurements are summarized in Appendix \ref{appdx:flux}.
Some details are given as follows.

\paragraph{DoAr~24Ea and DoAr~24Eb} These two binary components were both significantly detected in our SMA images (Appendix \ref{appdx:image}). With a 2$''$ angular separation, the images of them are confused in our SMA observations at $\lesssim$230 GHz, which makes the Gaussian fitting uncertain. Since the radii of these two binary components are $\lesssim$0\farcs1 (Table \ref{tab:protostars}), which cannot be resolved by our SMA observations (Section \ref{sub:imaging}), we treated these two binary components as distinct point sources. To derive their flux densities, we simultaneously fit two point sources to the observations on them at all frequencies. The free parameters include the flux densities (flux), x-offsets (dx), and y-offsets (dy) of the two point source models. 
The prior function of flux is uniform within the range of 0–2 Jy and is zero elsewhere. 
The prior functions of dx and dy are Gaussian functions with standard deviations equal to the synthesized beam FWHM, which are centered at the pixels of the source centers that we determined from the self-calibrated 254--318 GHz image (Section \ref{sub:imaging}).

\paragraph{IRS~37, IRS~39, and IRS~41} The source IRS~39 was not detected in the image domain at most of the observing frequencies (Table \ref{tab:obs_summary}) and thus was not considered in our visibility fittings. The 1\farcs92 separated companion of IRS~37 (Table \ref{tab:protostars}) was not detected in our SMA observations. Therefore, we treated IRS~37 as an isolated single source. The primary beams of our SMA observations on IRS~37, IRS~39, and IRS~41 overlapped. The primary beam FWHMs and the degrees of overlap depend on the observing frequency. For each of the two sources IRS~37 and IRS~41,  we jointly fit the observations that covered the sources within the primary beam FWHMs with circular Gaussian models. For example, when the observations at a certain frequency (e.g., 199 GHz) covered both IRS~37 and IRS~41 in two overlapping primary beams, we simultaneously fit the observations of these two antenna pointings with two circular Gaussian models.  
The free parameters of each circular Gaussian model include flux density (flux), x-offset (dx), y-offset (dy), and full width at half maximum (FWHM). 
When evaluating the likelihood function (i.e., when comparing models with the observed complex visibilities), the values of flux were implemented with the primary beam attenuation at the location of the intensity peaks. 
The prior functions for flux, dx, and dy were similar to those for the point source models we used in the fittings for DoAr~24Ea and Eb. 
The prior function for FWHM was wide but not flat, which is a Gaussian function centered at 0\farcs24 with a standard deviation equal to 4\farcs7. 
In addition, we rejected the MCMC samplers that walked to FHWM$<$0$''$.

\paragraph{Other sources in Table \ref{tab:protostars}}
Our SMA observations did not detect the companions of DoAr~16, YLW~8, and IRS~51 (Table \ref{tab:protostars}).
The 0\farcs43 separated binary components of HBC~266 (Table \ref{tab:protostars}) cannot be separated in our SMA observations.
Thus, we treated DoAr~16, YLW~8, IRS~51, HBC~266, and the other sources in Table \ref{tab:protostars} (except for DoAr~24Ea, DoAr~24Eb, IRS~37, IRS~39, and IRS~41) as isolated single sources.
For each of these isolated single sources, to derive flux densities, we fit the visibilities with a circular Gaussian model. 
The free parameters and prior functions are similar with those in the fittings for IRS~37 and IRS~41.

\begin{figure}
    \hspace{-0.3cm}
    \includegraphics[width=9cm]{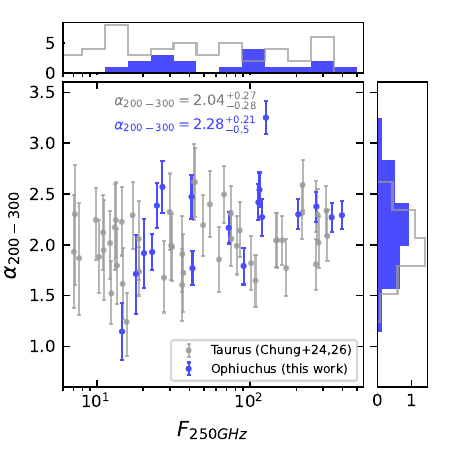}
    \caption{The 199--310 GHz spectral index ($\alpha_{\rm 200-310}$) versus 250 GHz flux density ($F_{\rm 250 GHz}$). The values of $F_{\rm 250 GHz}$ have been rescaled to a 140 pc distance, according to the distances of the target sources listed in Table \ref{tab:protostars}.
    The measurements for Taurus sources were quoted from \citet{Chung2024ApJS..273...29C} and  \citet{Chung2026arXiv260805445C}. For the faint Taurus sources quoted from \citet{Chung2026arXiv260805445C}, we approximated $\alpha_{\rm 200-300}$ by $\alpha_{\rm 198-358}$ since their observations did not cover the 255--330 GHz frequency. The histograms on the right have been normalized such that the area under each of them is 1.0.
    From the sample of \citet{Chung2024ApJS..273...29C}, we excluded the extended disks, DL~Tau, CI~Tau, GM~Tau, AB~Aur, DM~Tau, AA~Tau, and GO~Tau that have relatively large spectral indices, and excluded the source CoKu~Tau~1 and FY~Tau due to the poor data quality.
    }
    \label{fig:Fvsalpha}
\end{figure}

\subsection{Fitting spectral indices}\label{sub:SED_fitting}
We derived the $\sim$200--300 GHz spectral indices ($\alpha_{\rm 200-300}$) from our SMA observations on the selected Class II disks in the Ophiuchus molecular clouds, and from the previously published SMA observations on the Class II disks in the Taurus-Auriga region ( \citealt{Chung2024ApJS..273...29C,Chung2026arXiv260805445C}).

\paragraph{Ophiuchus sources}
For each source, our flux density measurements (Section \ref{sub:flux_density}) yielded 8 independent samples over the frequency range of [193, 318] GHz.
To derive spectral indices ($\alpha$), we fit the 8 flux density measurements of each source with a power-law using the MCMC method, which was implemented with the \texttt{emcee} package (\citealt{Foreman-Mackey2013PASP}). 
The free parameters include the flux density at 200 GHz frequency ($F_{\rm 200 GHz}$) and $\alpha$.
We assumed a flat prior function for $F_{\rm 200 GHz}$ and rejected the MCMC samplers that walked to $F_{\rm 200 GHz}<$0 Jy.
The prior function of $\alpha$ is uniform in the range of 1.0--5.0 and is zero elsewhere.

\paragraph{Taurus-Auriga sources}
\citet{Chung2024ApJS..273...29C} published the uniformly sampled 12 flux density measurements over the [198, 415.5] GHz frequency range for 47 Class II disks in the Taurus-Auriga region. 
We quoted their flux density measurements in the frequency range of [199, 310] GHz, excluding the exceptionally extended sources DL~Tau, CI~Tau, GM~Tau, AB~Aur, DM~Tau, AA~Tau, and GO~Tau that have relatively large spectral indices, and excluding the source CoKu~Tau~1 that was subject to poor data quality. 
We also excluded the source FY~Tau, in which the flux density at 310 GHz appears biased, which is likely due to measurement errors (see Figure 7 of \citealt{Chung2024ApJS..273...29C}).
We re-fit the $\alpha_{\rm 200-300}$ values using the procedure mentioned above. 
In addition, \citet{Chung2026arXiv260805445C} extended the SMA survey to 10 fainter ($F_{\rm 225 GHz}\lesssim$10 mJy) Class II disks in the Taurus-Auriga region. 
In this survey for fainter sources, they sampled the flux densities at [198, 220, 227, 249, 336, 358, 407.5] GHz frequencies.
We approximated the $\alpha_{\rm 200-300}$ of these 10 fainter sources by fitting the observations in the frequency range of 198--358 GHz using the procedure mentioned above.
We do not think that this will lead to a noticeable systematic bias in the comparison with our present observations of the Ophiuchus sources, since the difference in observing frequencies is small, and since the flux densities and $\alpha_{\rm 200-300}$ of these 10 fainter sources have large error bars. 
In total, the Taurus-Auriga sample we are comparing with comprises 48 spatially compact Class II disks. 

\begin{deluxetable*}{lccccrr}
\tabletypesize{\scriptsize}
\tablecolumns{7}
\tablewidth{0pt}
\tablecaption{Comparison of the 200--300 GHz and 91--345 GHz spectral indices
\label{tab:alpha91340}}
\tablehead{
\colhead{Source} &
\colhead{$\nu_{\rm SMA}$}\tablenotemark{a} &
\colhead{$F_{\rm SMA}$}\tablenotemark{a} &
\colhead{$F_{\rm 91GHz}$}\tablenotemark{b} &
\colhead{$\alpha_{\rm 91-340}$} &
\colhead{$\alpha_{200-300}$} &
\colhead{$\Delta\alpha$} \\
\colhead{} &
\colhead{(GHz)} &
\colhead{(mJy)} &
\colhead{(mJy)} &
\colhead{} &
\colhead{} &
\colhead{}
}
\startdata
GSS 39 &
340.0 &
$663\pm3$ &
$38.7\pm0.6$ &
$2.16\pm0.20$ &
$2.27\pm0.14$ &
$+0.11\pm0.25$ \\
DoAr 25 &
346.6 &
$563\pm3$ &
$25.0\pm0.6$ &
$2.33\pm0.20$ &
$2.38^{+0.15}_{-0.14}$ &
$+0.05\pm0.25$ \\
YLW 8 / SR 21 &
346.6 &
$371\pm2$ &
$4.2\pm0.4$ &
$3.35\pm0.21$ &
$3.25^{+0.17}_{-0.16}$ &
$-0.10\pm0.27$ \\
VSSG 1 &
340.0 &
$275\pm2$ &
$7.3\pm0.5$ &
$2.75\pm0.21$ &
$2.54\pm0.17$ &
$-0.21\pm0.27$ \\
WSB 60 &
339.5 &
$255\pm3$ &
$15.3\pm0.5$ &
$2.14\pm0.21$ &
$2.27\pm0.18$ &
$+0.13\pm0.27$ \\
DoAr 44 &
339.5 &
$229\pm3$ &
$10.4\pm0.5$ &
$2.35\pm0.21$ &
$2.42\pm0.18$ &
$+0.07\pm0.28$ \\
\enddata
\tablenotetext{a}{Quoted from \citet{Andrews2009ApJ...700.1502A}.}\vspace{-0.2cm} 
\tablenotetext{b}{Quoted from \citet{Ricci2010}.}\vspace{-0.2cm}
\tablenotetext{c}{Defined as $\Delta\alpha=\alpha_{200-300}-\alpha_{\rm 91-340}$.}\vspace{-0.2cm}
\tablecomments{
The uncertainties quoted for the flux densities are the statistical/rms uncertainties that do not include the absolute flux calibration uncertainties.
In estimating the uncertainties of $\alpha_{\rm 91-340}$, we combined the tabulated statistical uncertainties with absolute flux calibration uncertainties of 10\% for the SMA measurements (\citealt{Andrews2009ApJ...700.1502A}) and 25\% for the ATCA measurements (\citealt{Ricci2010}).
}
\end{deluxetable*}

\section{Results} \label{sec:results}
Figure \ref{fig:SED} and \ref{fig:SED_binary} show the flux density measurements for the 18 Class II disks covered by our SMA observations (Section \ref{sec:observation}) and the results of our spectral index fittings (Section \ref{sub:SED_fitting}).
In the frequency range of 200--250 GHz, our flux density measurements are reasonably consistent with the ALMA observations reported in \citet{Andrews2018ApJ...869L..41A,Garufi2020A&A...633A..82G,Testi2022A&A...663A..98T,Michel2023AJ....166..184M,Shoshi2025PASJ...77..572S} and the SMA observations reported in \citet{Andrews2009ApJ...700.1502A,Andrews2010ApJ...723.1241A,Tripathi2017ApJ...845...44T}.

At $\sim$350 GHz, the power-laws fitting our SMA observations are consistent with the previous SMA and ALMA measurements of \citet{Andrews2008ApJ...678L.133A,Andrews2009ApJ...700.1502A,Andrews2010ApJ...723.1241A}, \citet{Tripathi2017ApJ...845...44T}, \citet{Garufi2020A&A...633A..82G}, and \citet{Michel2023AJ....166..184M} to $\pm$1-$\sigma$. 
The ALMA measurements of 345 and 410 GHz flux densities reported in \citet{Cox2017ApJ...851...83C} and \citet{Dasgupta2025ApJ...981L...4D} are either very consistent with the power-laws we derived (e.g. WSB~60, DoAr~44, AS~206, etc) or are (both) systematically lower than the power-laws we derived (e.g. WSB~31, DoAr~25, GSS~39, etc).
As WSB~31, DoAr~25, and GSS~39 are the most spatially extended disks in our sample (Table \ref{tab:protostars}), we attribute the discrepancy between our power-law fits and these measurements of \citet{Cox2017ApJ...851...83C} and \citet{Dasgupta2025ApJ...981L...4D} to the missing short-spacing in the observations of \citet{Cox2017ApJ...851...83C} and \citet{Dasgupta2025ApJ...981L...4D}.
The single-dish measurements of \citet{Andrews2007ApJ...671.1800A} are largely offset from these power-laws, which may be due to calibration errors.
These comparisons, in particular, the excellent consistency between our best-fit power-laws and the measurements of \citet{Andrews2008ApJ...678L.133A,Andrews2009ApJ...700.1502A,Andrews2010ApJ...723.1241A,Tripathi2017ApJ...845...44T}, provide invaluable clarification for the accuracy of flux densities in the literature. 

Only one source, YLW~8, which has a large inner cavity (\citealt{Yang2023ApJ...948..110Y}), shows a spectral index $\alpha_{\rm 200-300}>3$; and only one source, IRS~37, shows $\alpha_{\rm 200-300}<1.5$.
The relatively low $\alpha_{\rm 200-300}$ value of IRS~37 may indicate that it is confused by free-free emission (e.g. \citealt{Liu2014ApJ...780..155L,Coutens2019A&A...631A..58C}).

Figure~\ref{fig:Fvsalpha} shows the relation between the 250 GHz flux density ($F_{\rm 250 GHz}$) and the spectral index $\alpha_{200-300}$.
The value of $F_{\rm 250GHz}$ was interpolated from the results of our spectral-index fitting and have been rescaled to a common distance of 140 pc according to the source distances in Table \ref{tab:protostars}.
Among the Ophiuchus sources, the median $\alpha_{\rm200-300}$ is 2.28; the positive and negative standard 1-$\sigma$ of the $\alpha_{\rm200-300}$ distribution are 0.21 and 0.5 (Figure \ref{fig:Fvsalpha}), which is determined as the 16th and 84th percentiles of the posteriors in a Bayesian inference.
This $\alpha_{\rm 200-300}$ distribution is consistent within $1\sigma$ with that measured for the 48 compact Taurus--Auriga disks (Section \ref{sub:SED_fitting}), which have $\alpha_{\rm 200-300}=2.04\pm0.3$.
The $\alpha_{200-300}$ distribution of the Ophiuchus sample shows a small skewness\footnote{For a random variable $X$, skewness is the third standard moment $\frac{E[(X-\mu)^3]}{\sigma^3}$, where $E[\cdot]$ denotes the expectation value, $\mu$ is the mean of the distribution and $\sigma$ is the standard deviation.} of $-$0.11.

We used the Kolmogorov–Smirnov (KS) test and the non-parametric Mann–Whitney U test to examine the null hypothesis that the $\alpha_{\rm 200-300}$ distributions of the Ophiuchus and Taurus-Auriga samples are not distinguishable. 
This is implemented using the \texttt{ks\_2samp} and \texttt{mannwhitneyu} functions of the scipy package (\citealt{2020SciPy-NMeth}).
In the KS test, the median $p$ value is 0.21, which cannot reject the null hypothesis.
To assess the robustness of this statement, we re-evaluated the $p$ values of 5000 random realizations of the actual $\alpha_{\rm 200-300}$ values assuming that the errors of $\alpha_{\rm 200-300}$ followed Gaussian distributions.
We found that the fraction of $p<$0.05 is 0.144.
In the Mann–Whitney U test, the median $p$ value is 0.21; the fraction of $p<$0.05 is 0.099.
With these, we concluded that the Ophiuchus and Taurus-Auriga samples are consistent with the same underlying $\alpha_{\rm 200-300}$ distribution. 

\begin{figure}
    \hspace{-0.3cm}
    \includegraphics[width=9cm]{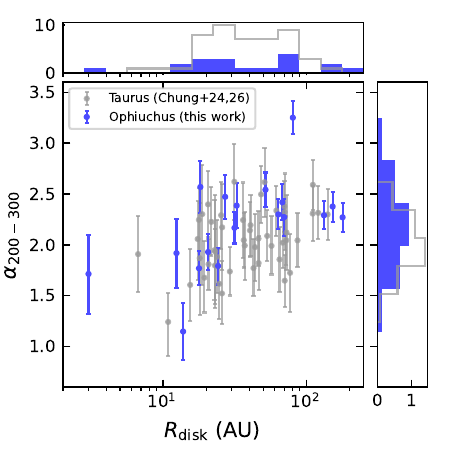}
    \caption{The 199--310 GHz spectral index ($\alpha_{\rm 200-300}$) versus disk radius ($R_{\rm 95\%}$). The histograms on the right have been normalized such that the area under each of them is 1.0.}
    \label{fig:Rvsalpha}
\end{figure}
 
For the Ophiuchus and Taurus–Auriga samples, the Pearson correlation coefficients between $F_{\rm 250GHz}$ and $\alpha_{200-300}$ are 0.30 ($p = 0.22$) and 0.23 ($p = 0.11$), respectively (Figure~\ref{fig:Fvsalpha}).
Jointly analyzing the Ophiuchus and Taurus samples yielded a Pearson correlation coefficient of 0.27 ($p=0.02$).
If we exclude the YLW~8 disk that has a exceptionally high $\alpha_{200-300}$ from the Ophiuchus sample, the Pearson correlation of the joint sample becomes 0.28 ($p=0.02$).
If Ophiuchus and Taurus-Auriga samples are indeed consistent with the same underlying $\alpha_{\rm 200-300}$ distribution, then $F_{\rm 250GHz}$ and $\alpha_{200-300}$ might be correlated. 

Figure \ref{fig:Rvsalpha} shows the relations between $\alpha_{200-300}$ and the effective radius of disk ($R_{\rm 95\%}$), where $R_{\rm 95\%}$ is the radius of a circular region that encloses 95\% of the total flux density (Appendix \ref{appdx:radii}).
The measurements from the Ophiuchus and Taurus samples present very similar distributions in this plot.
Among the Ophiuchus and Taurus samples, the Pearson correlation coefficients are 0.40 ($p=0.099$) and $-0.012$ ($p=$0.93), respectively.
Jointly analyzing the Ophiuchus and Taurus samples yielded the Pearson correlation coefficient of $-0.02$ ($p=$0.86).
If we exclude the YLW~8 disk that has a exceptionally high $\alpha_{200-300}$ from the Ophiuchus sample, the Pearson correlation of the joint sample becomes $-0.017$ ($p=$0.89).
Out of these analyses, we do not detect a convincing correlation between $\alpha_{200-300}$ and $R_{\rm 95\%}$.

To examine whether the spectral indices measured over 200–300 GHz differ from those measured over a substantially broader wavelength range, we considered the six sources in our sample that were observed interferometrically with the SMA by \citet{Andrews2009ApJ...700.1502A} at $\sim$340 GHz and with ATCA by \citet{Ricci2010} at $\sim$91 GHz (Table \ref{tab:alpha91340}). We derived a two-point broad-band spectral index, $\alpha_{\rm 91-340}$, assuming \(F_\nu\propto\nu^\alpha\). We find that $\alpha_{\rm 91-340}$ is consistent with our $\alpha_{\rm 200-300}$ measurements for all six sources within 1-$\sigma$. The differences $\alpha_{\rm 200-300}-\alpha_{\rm 91-340}$ have a mean of 0.01 and a median of 0.06, indicating no evidence for a systematic difference between the local 200--300 GHz spectral slopes and the broad-band 91--340 GHz slopes.
This result is similar to that observed for some Class II disks in the Taurus--Auriga star-forming region (\citealt{Chung2025ApJS..277...45C}), which exhibit an approximately constant spectral slope over the frequency range of 90--400 GHz and a steeper slope at $\sim$50--90 GHz frequencies.

\section{Discussion and summary} \label{sec:discussion}

We performed SMA observations on 18 selected Class II disks in the Ophiuchus molecular cloud.
The flux densities of our selected sources are in the range of 10--500 mJy at the frequency of 250 GHz. 
Our observations yield 8 independent measurements of flux densities over the frequency range of 199--310 GHz. 
Based on these measurements, the derived spectral indices $\alpha_{\rm 200-300}$ are 2.3$^{+0.2}_{-0.5}$ (Figure \ref{fig:Fvsalpha}).

We estimated lower limits on the dust masses using uniform-slab dust emission models that incorporate both dust absorption and scattering.
We adopted the standard DSHARP dust opacities (\citealt{Birnstiel2018ApJ...869L..45B}).
Owing to the known degeneracy between dust grain size (or the dust opacity spectral index) and dust temperature, we followed the approach of \citet{Chung2024ApJS..273...29C} and considered two representative combinations of dust temperature and maximum grain size, $(T_{\rm dust},a_{\rm max})=$(24~K, 3~mm) and (100~K, 0.05~mm).
For each case, we generated model flux densities at our eight observing frequencies (Table~\ref{tab:obs_summary}) while varying the dust surface density, $\Sigma_{\rm dust}$, from 0.1 to 5.0~g~cm$^{-2}$.
We measured the model spectral index, $\alpha_{200-300}$, by fitting a power law to the synthetic flux densities.
For each source, we identified the minimum surface density, $\Sigma_{\rm dust,min}$, at which the model spectral index became lower than the observed value.
If this condition was not reached within the explored range, we adopted $\Sigma_{\rm dust,min}=5.0$~g~cm$^{-2}$, in which case the inferred surface density represents a lower limit.
Holding $\Sigma_{\rm dust}$ fixed at $\Sigma_{\rm dust,min}$, we then varied the solid angle of the emitting region until the model 250~GHz flux density matched the observed $F_{\rm 250\,GHz}$.
The corresponding solid angle, $\Omega_{\rm min}$, was combined with the source distance to estimate the dust mass as $M_{\rm dust}=\Sigma_{\rm dust,min}\Omega_{\rm min}d^2$.
We performed this procedure independently for the two $(T_{\rm dust},a_{\rm max})$ combinations to bracket the dust masses inferred for these representative dust properties.

\begin{figure}
  \hspace{-0.3cm}
  \includegraphics[width=9cm]{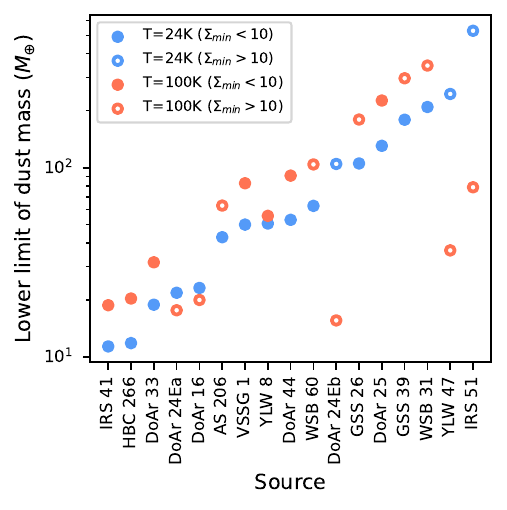}
  \vspace{-0.8cm}
  \caption{
  Lower limits on the dust masses derived from the observed $\alpha_{200-300}$ and $F_{\rm 250\,GHz}$ (for details, see Section \ref{sec:discussion}). Blue and red symbols show the estimates obtained by adopting $(T_{\rm dust},a_{\rm max})=$(24~K, 3~mm) and (100~K, 0.05~mm), respectively. Open circles indicate sources for which the model spectral index does not become lower than the observed $\alpha_{200-300}$ within the explored range of $\Sigma_{\rm dust}$; their inferred dust masses therefore correspond to the adopted upper boundary of $\Sigma_{\rm dust}$ and should be interpreted as lower limits. Filled circles indicate sources for which a minimum $\Sigma_{\rm dust}$ is identified within the explored range.
  }
  \label{fig:mass}
\end{figure}

Figure~\ref{fig:mass} summarizes the results of our dust mass estimates.
We exclude IRS~37 from the estimates because its exceptionally low spectral index (Figure \ref{fig:SED}) indicates a significant contribution from free--free emission.
We emphasize that both $\Sigma_{\rm dust,min}$ and $M_{\rm dust}$ inferred with this method should be regarded as lower limits.
This is because the disk-averaged $\alpha_{200-300}$ reflects a mixture of optically thick and optically thin dust emission.
The continuum emission does not directly constrain the amount of dust hidden in optically thick regions, and therefore potentially some amounts of dust mass can remain undetected.

For several spatially extended disks, such as GSS~26, DoAr~25, GSS~39, and WSB~31, a mean dust temperature of $\sim$24~K may be more realistic than 100~K (Figure~\ref{fig:mass}).
Even under this assumption, the estimated lower limits on the dust masses are $\gtrsim100~M_{\oplus}$ for a substantial fraction of our observed sources (Figure~\ref{fig:mass}).
These sources may preserve their a significant fraction of the primordial dust mass budget, which is not yet dispersed (e.g., due to inward migration or feedback) or being consumed in planet formation.
For other sources, the estimated lower limits on the dust masses lie in the range of $\gtrsim10$--100~$M_{\oplus}$ (Figure~\ref{fig:mass}).
If these lower limits are not substantially below their actual dust masses, a fraction of their primordial dust mass may already have been dispersed or incorporated into larger bodies during planet formation, as has been discussed in previous studies of younger sources (e.g., \citealt{Tychoniec2020}).
However, because our estimates provide only lower limits, we cannot exclude the possibility that all of the millimeter-luminous ($F_{\rm 250\,GHz}>10$~mJy when scaled to a distance of 140~pc), $\sim$1--2~Myr-old Class~II disks in our sample have largely preserved their primordial dust mass budgets.

Finally, given the current measurement uncertainties of $\alpha_{\rm 200-300}$, which were largely limited by thermal noise in the SMA observations, the data did not convincingly reveal the correlations between $\alpha_{\rm 200-300}$ and the $R_{\rm 95\%}$ disk radius (Figures \ref{fig:Fvsalpha}, \ref{fig:Rvsalpha}).
The $\alpha_{\rm 200-300}$ and the 250 GHz flux density ($F_{\rm 250GHz}$) may be positively correlated, with a Pearson correlation coefficient of 0.27 ($p=0.02$).
In other words, in the $F_{\rm 250GHz}$ range of the Ophiuchus and Taurus-Auriga disks we compared, there is no evidence that the fainter or smaller disks are optically thinner (e.g., due to loss of dust mass).

\begin{acknowledgments}
The National Radio Astronomy Observatory is a facility of the National Science Foundation operated under cooperative agreement by Associated Universities, Inc.
The Submillimeter Array is a joint project between the Smithsonian Astrophysical Observatory and the Academia Sinica Institute of Astronomy and Astrophysics and is funded by the Smithsonian Institution and the Academia Sinica.
We recognize that Maunakea is a culturally important site for the indigenous Hawaiian people; we are privileged to study the cosmos from its summit.
C.Y.L., C.Y.C., and H.B.L. are supported by the National Science and Technology Council (NSTC) of Taiwan (Grant Nos. 113-2112-M-110-022-MY3).
\end{acknowledgments}

\begin{contribution}

C.Y.C. prepared the proposal and the observing scripts, calibrated the data taken in 2023, and made preliminary development of the self-calibration procedure.
H.B.L. calibrated the data taken in 2022, performed self-calibration, and carried out visibility and SED fittings.
C.-Y.L. assisted collecting the previous measurements from the literature.
L.F. derived flux densities of several sources from the latest ALMA observations and made comparison. 
S.M.A assisted visibility fittings.
M.A.G. assisted the planning of observations and absolute flux calibrations.
M.W. assisted the gain-phase self-calibration.
All authors participated in the scientific discussion. 

\end{contribution}

%
\facilities{SMA, ALMA}

\software{
          astropy \citep{astropy2022ApJ...935..167A},  
          Numpy \citep{VanDerWalt2011}, 
          MIR IDL \citep{Qi2003cdsf.conf..393Q},
          Miriad \citep{Sault1995ASPC...77..433S},
          CASA \citep[][]{CASA2022PASP..134k4501C},
          emcee \citep{Foreman-Mackey2013PASP},
          scipy \citep{2020SciPy-NMeth},
          }


\appendix

\section{Image of target sources}\label{appdx:image}

\begin{figure*}
\hspace{-1.6cm}
  \begin{tabular}{ cccc }
  \includegraphics[width=4cm]{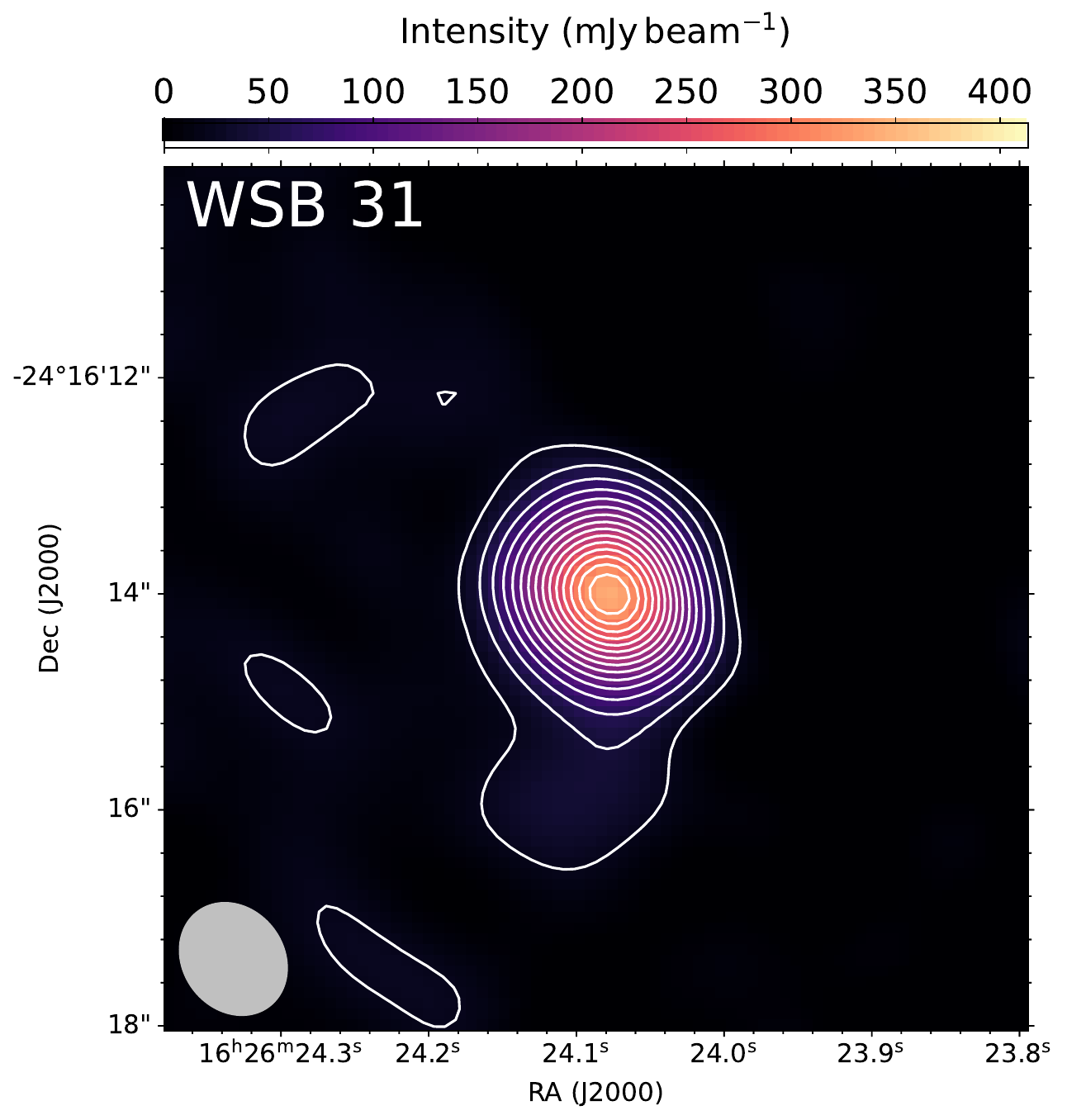} &
  \includegraphics[width=4cm]{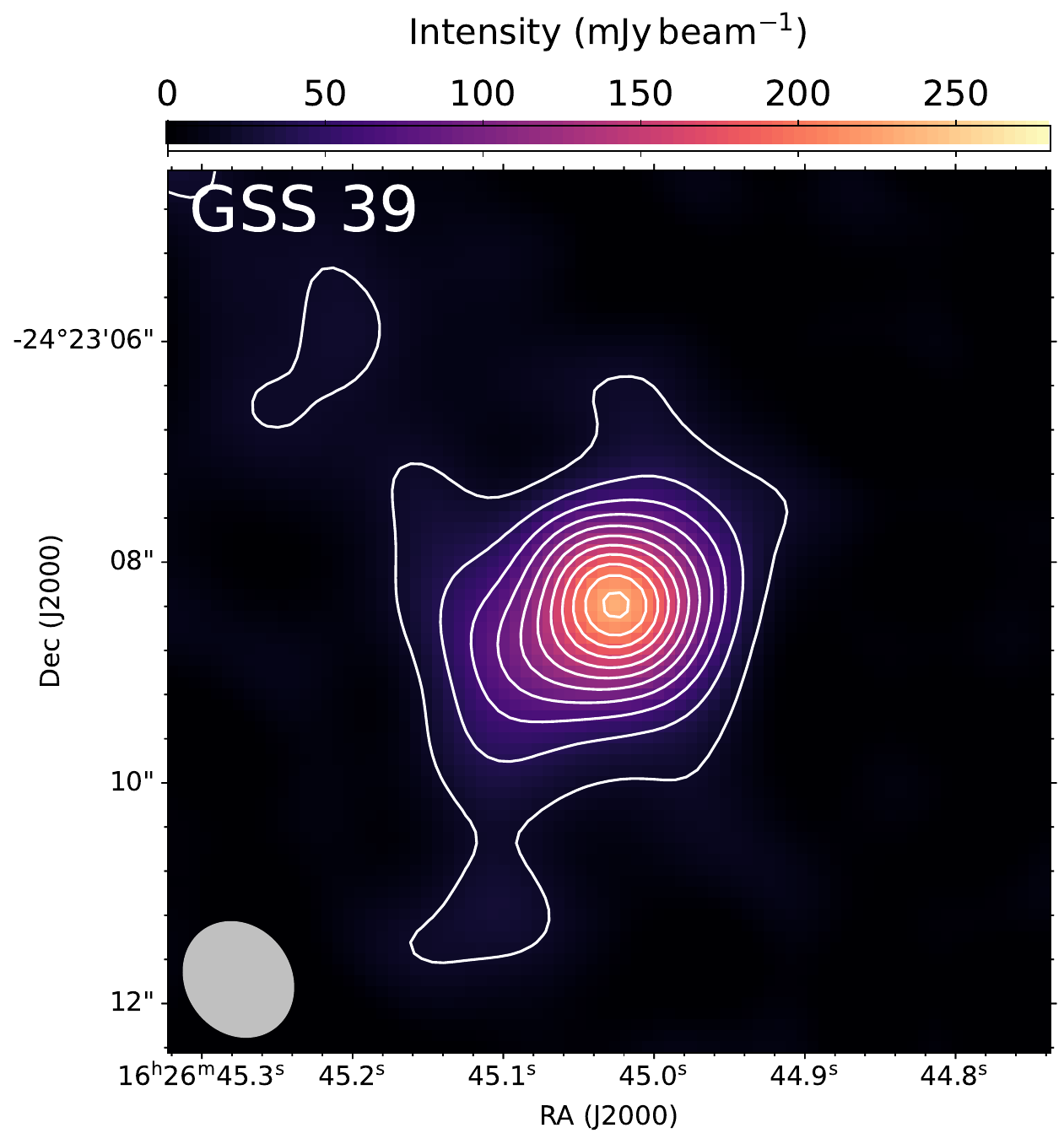} &
  \includegraphics[width=4cm]{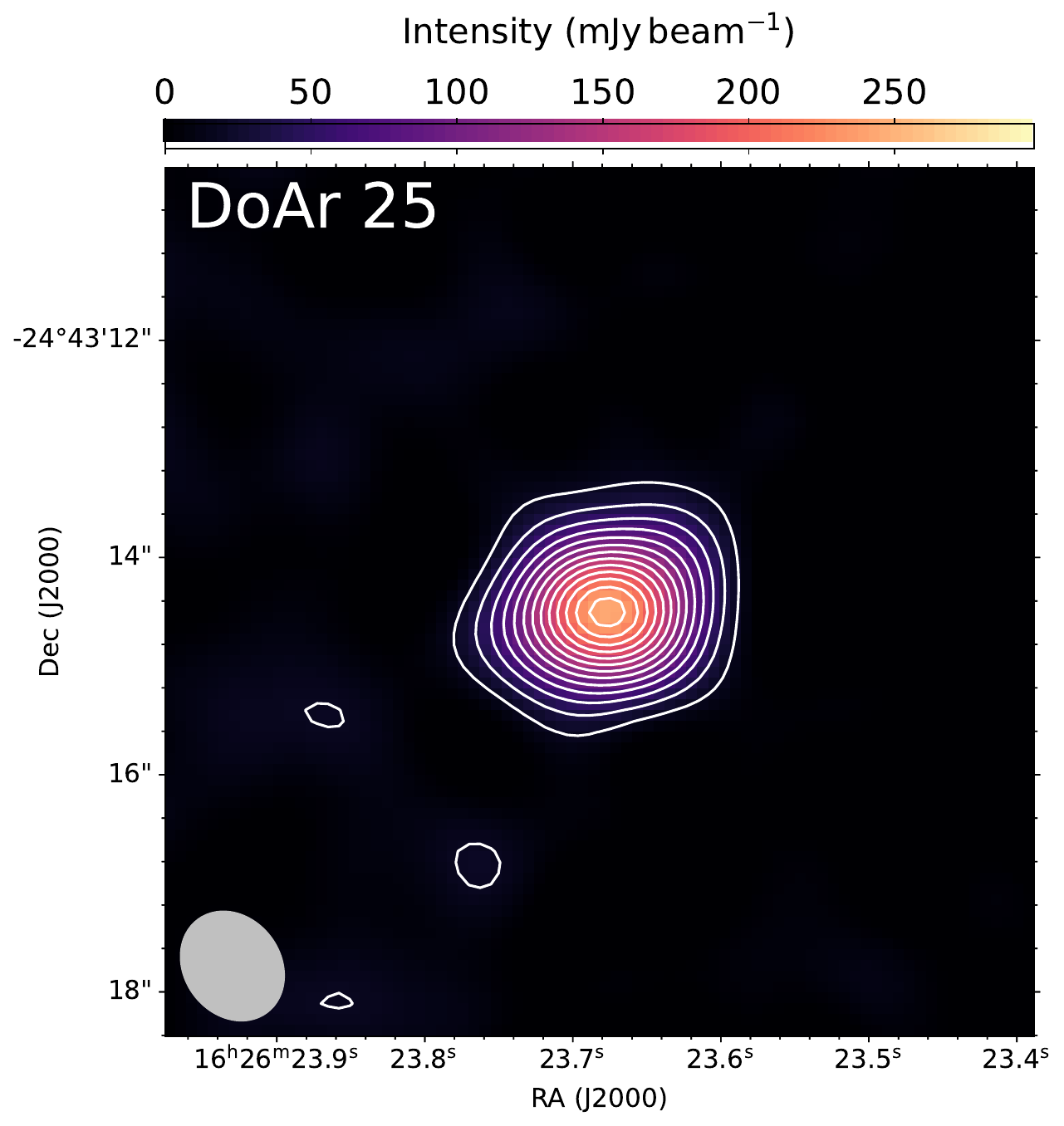} &
  \includegraphics[width=4cm]{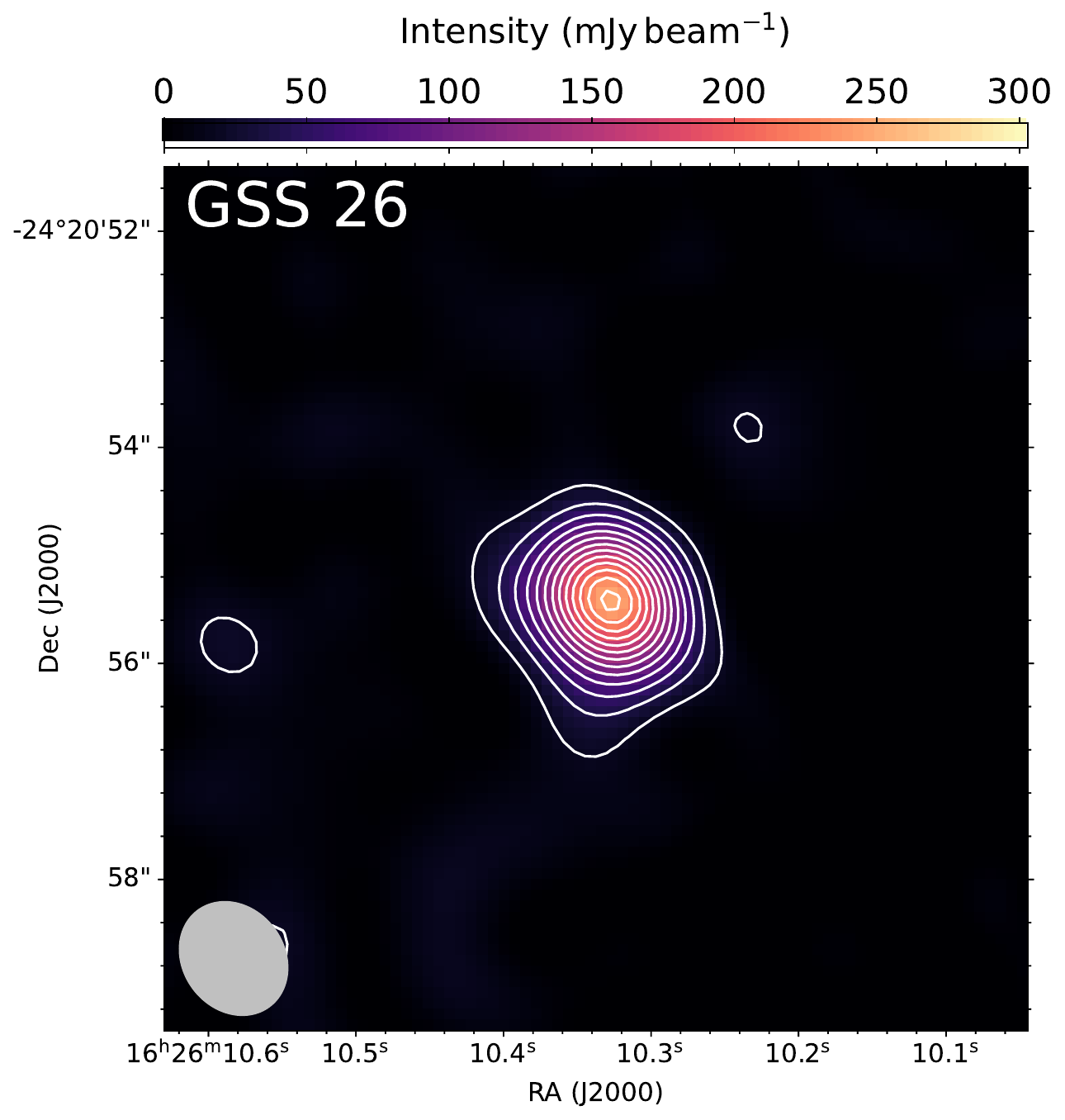}  \\
  \includegraphics[width=4cm]{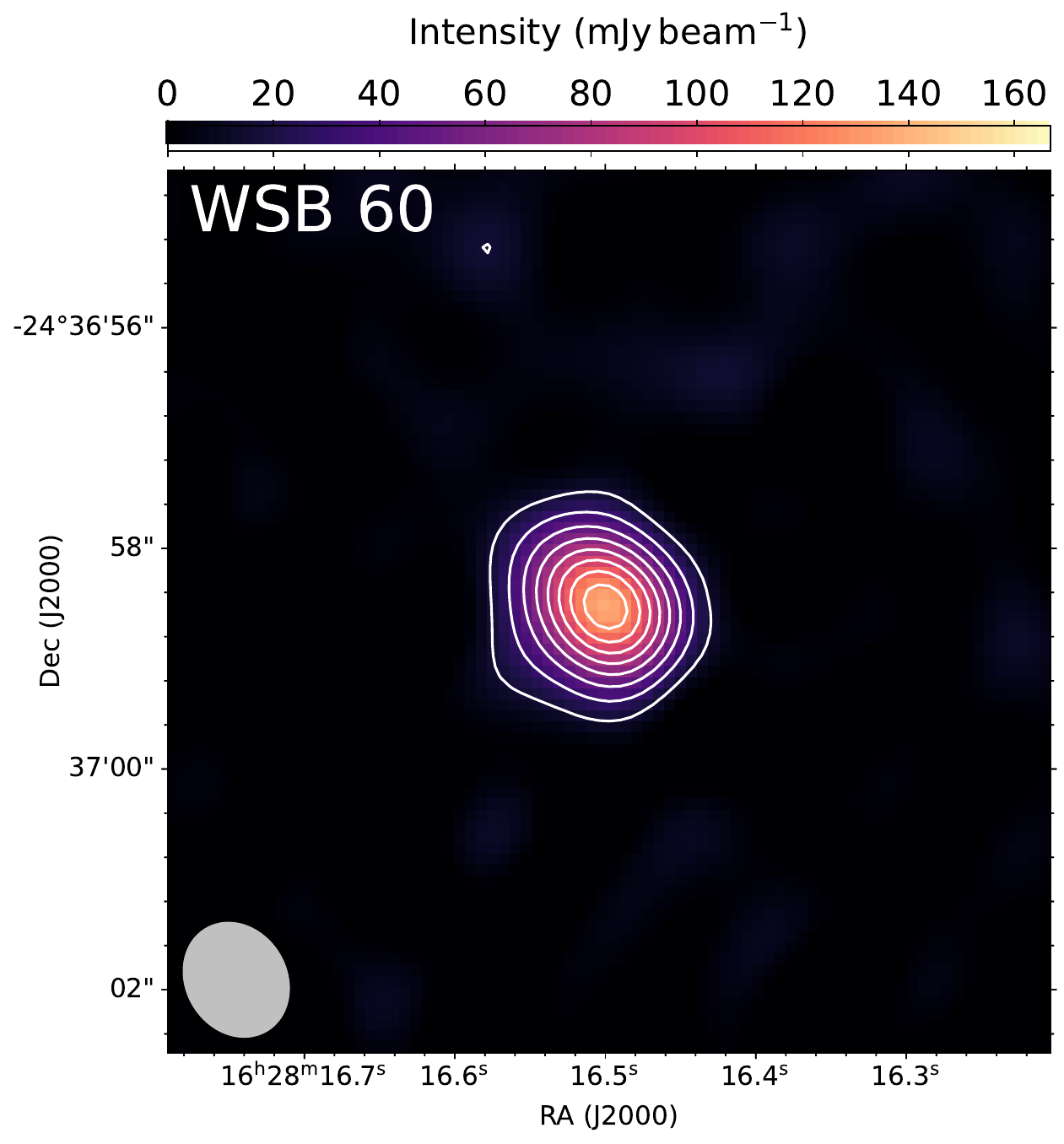} &
  \includegraphics[width=4cm]{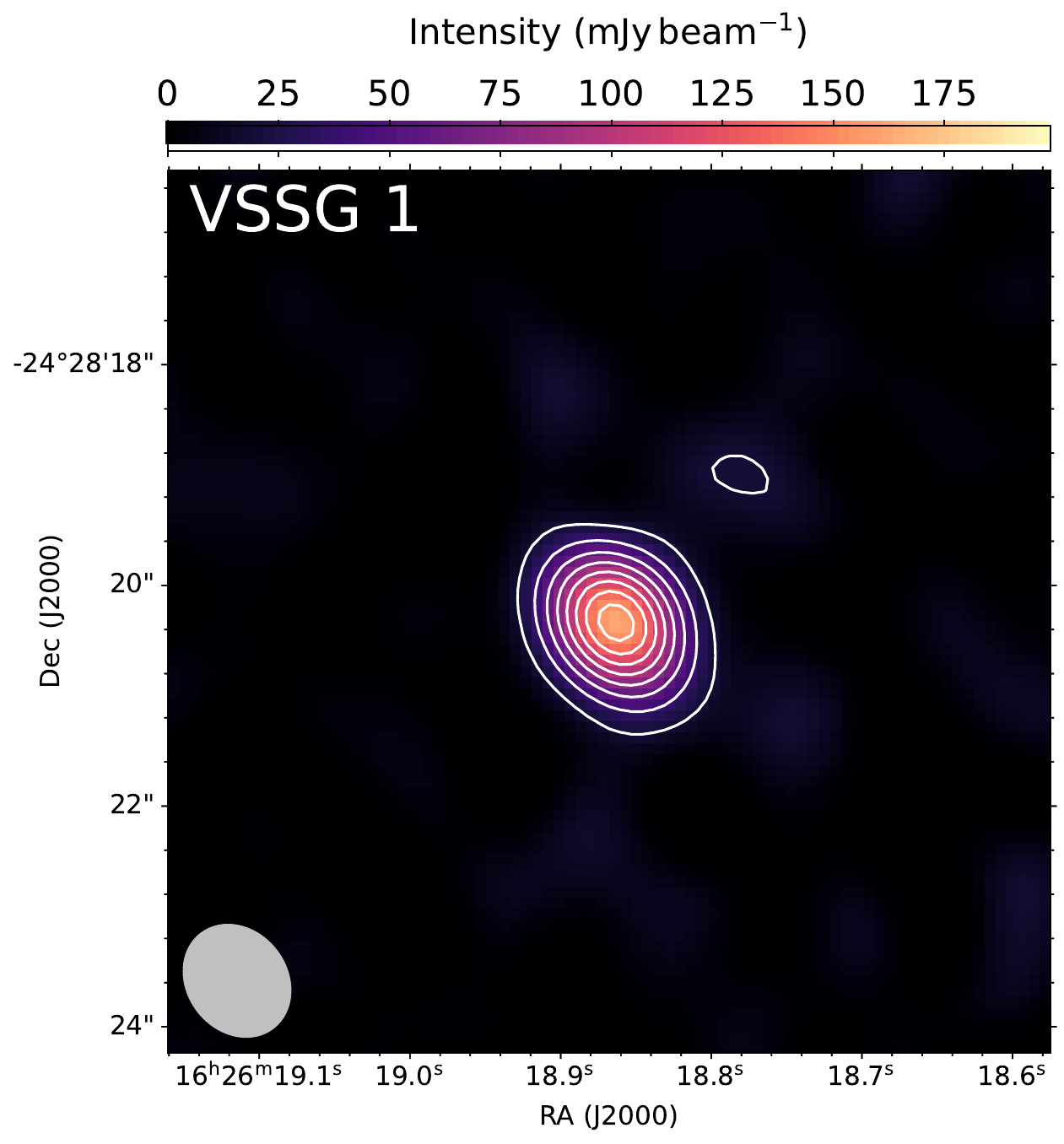} &
  \includegraphics[width=4cm]{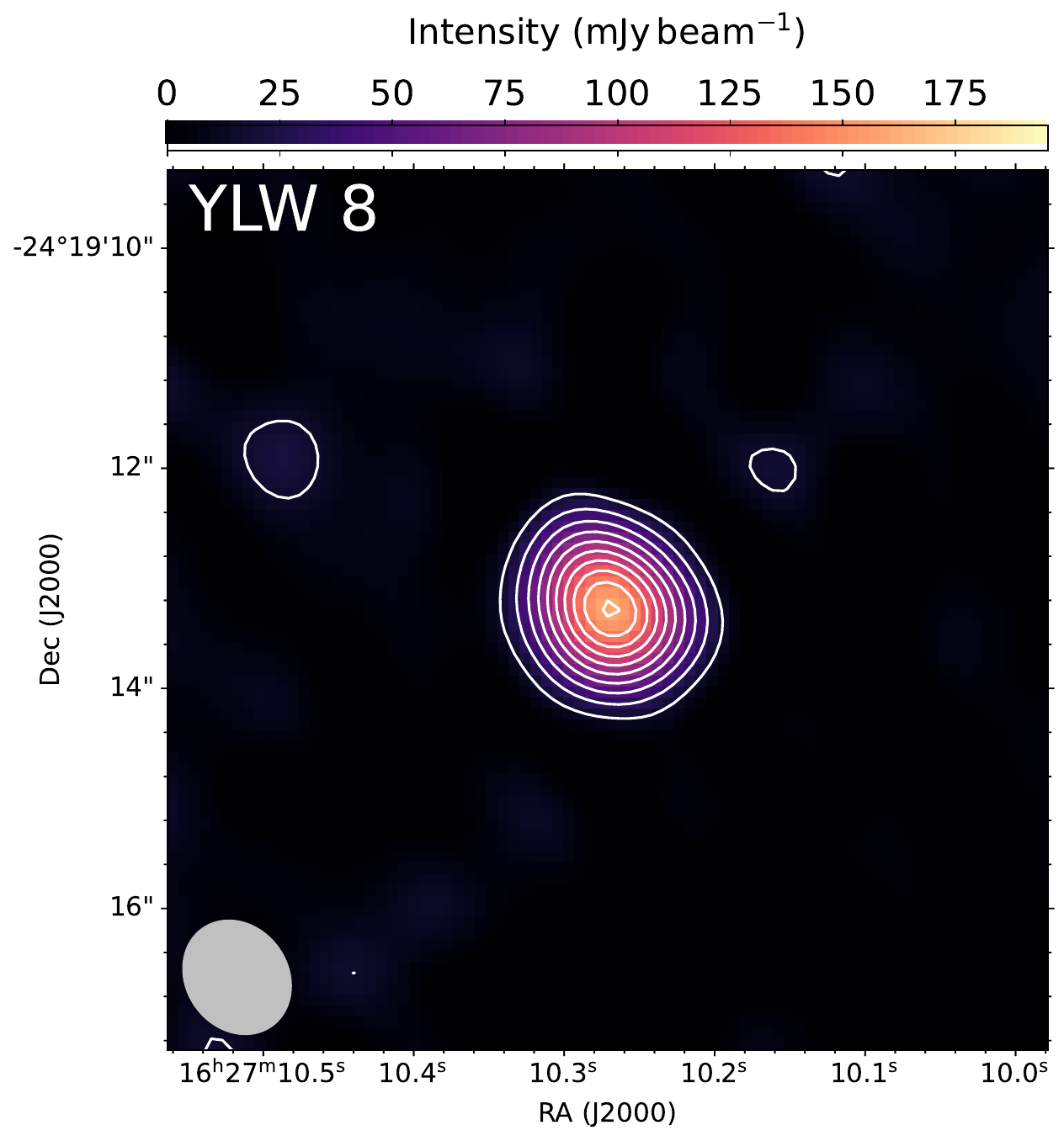} &
  \includegraphics[width=4cm]{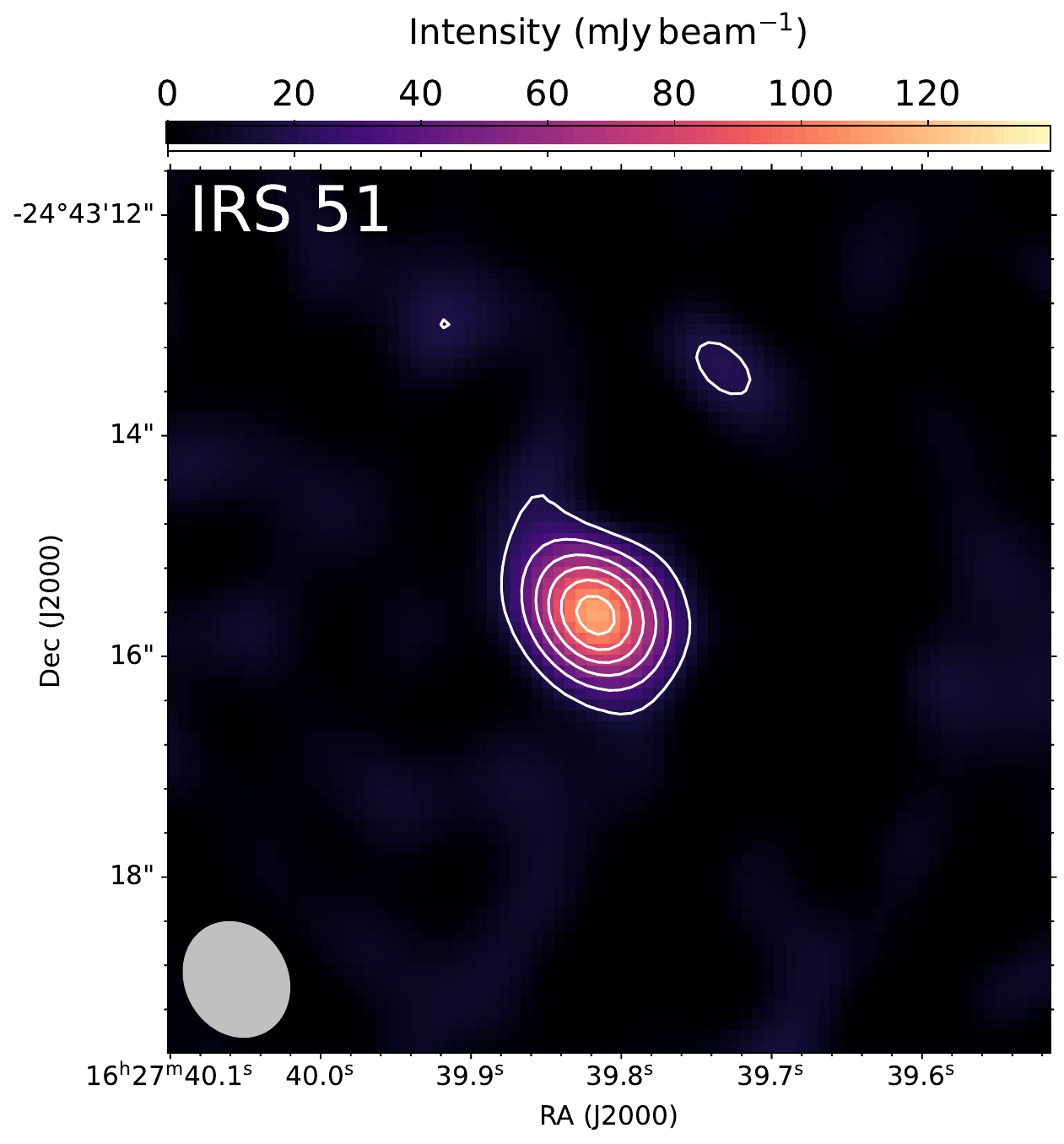} \\
  \includegraphics[width=4cm]{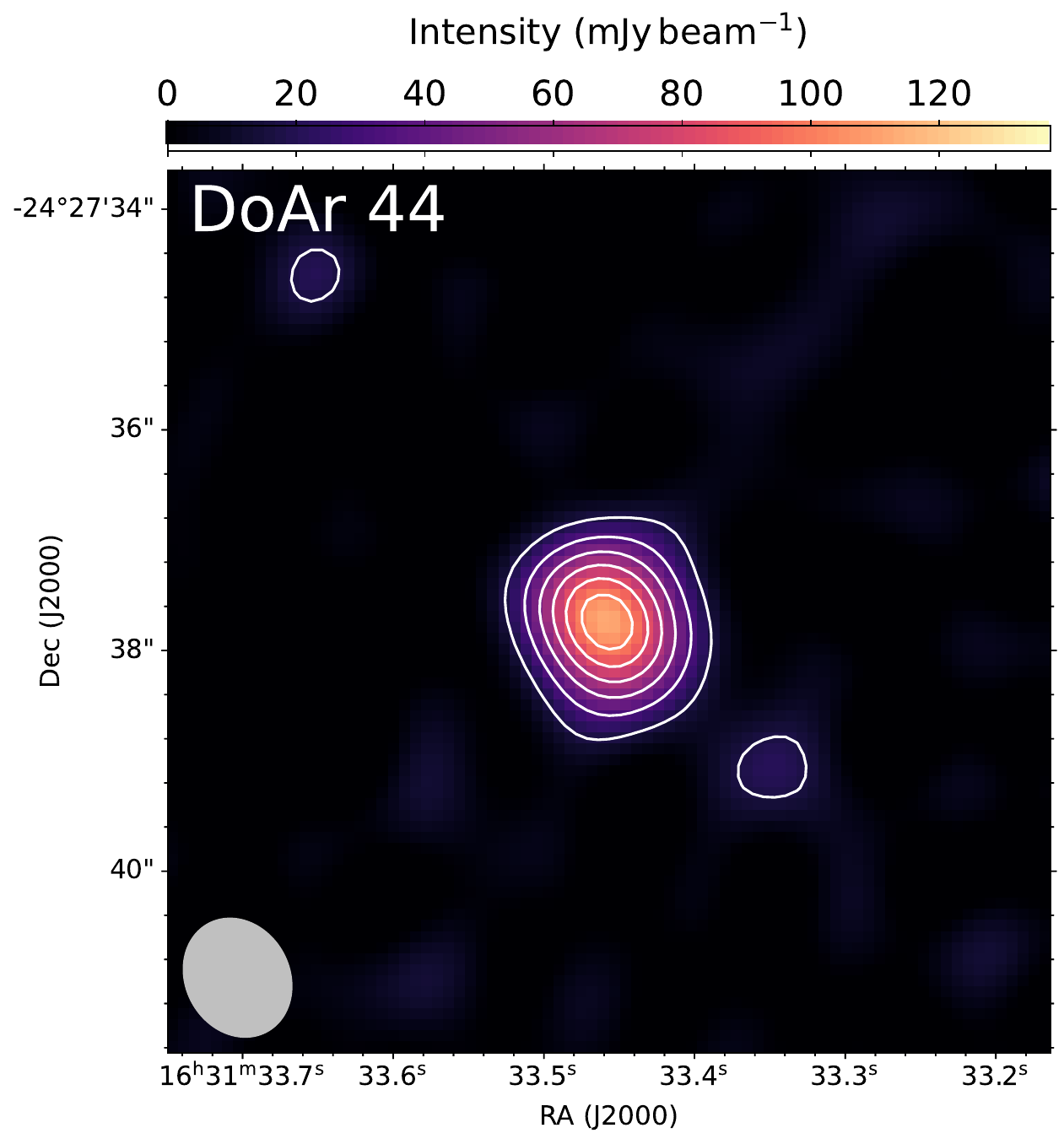} & 
  \includegraphics[width=4cm]{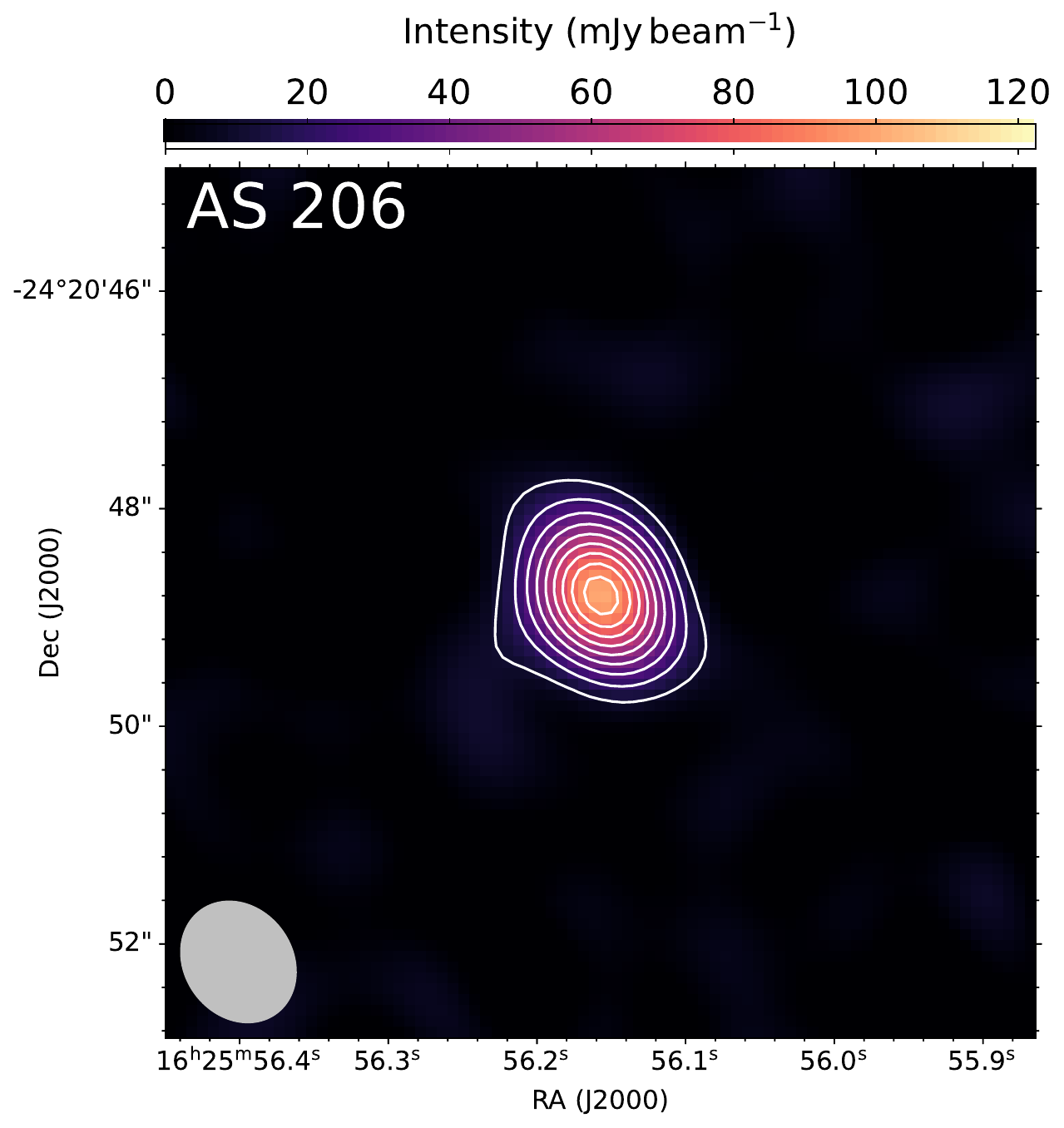} & 
  \includegraphics[width=4cm]{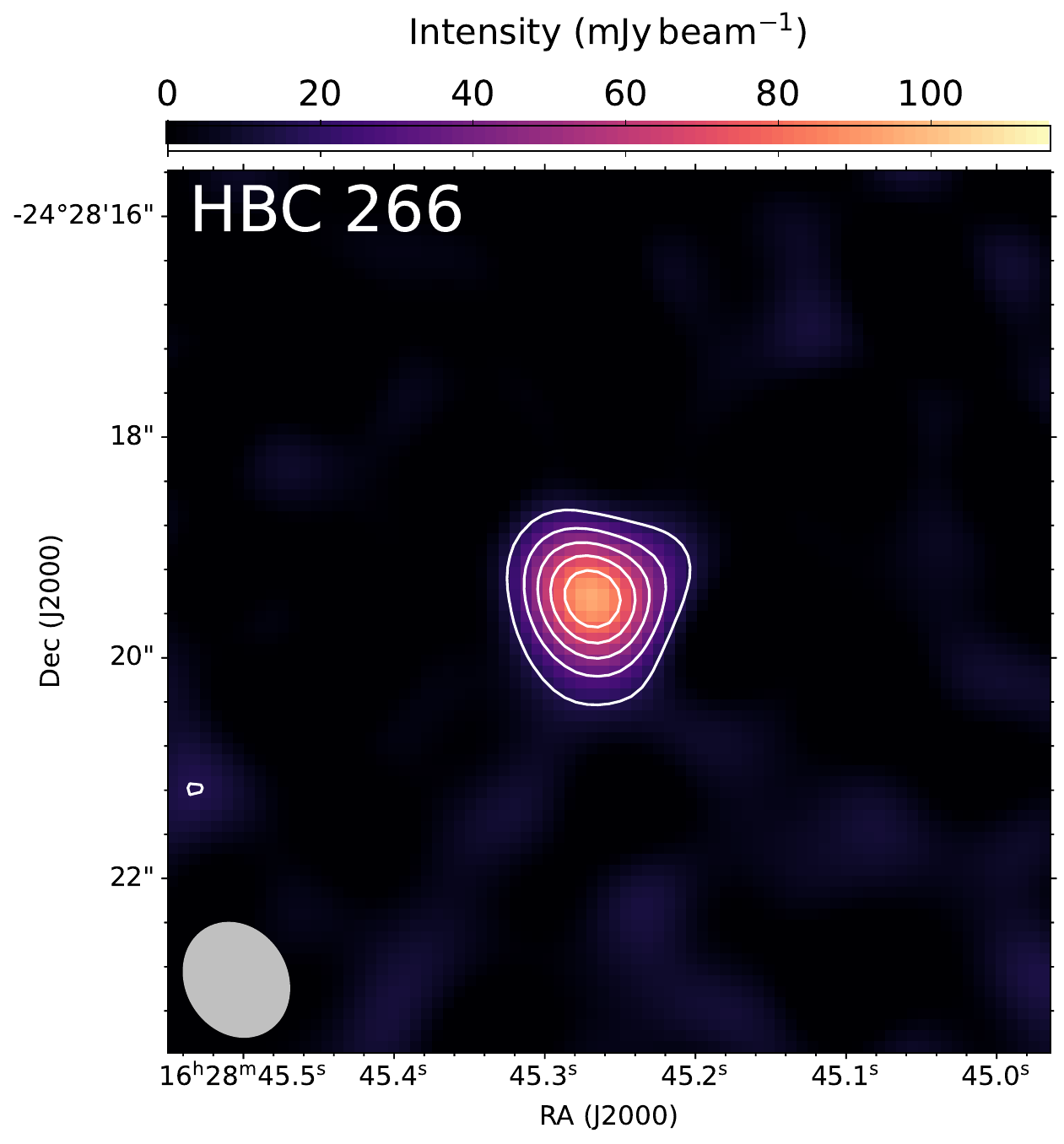}& 
  \includegraphics[width=4cm]{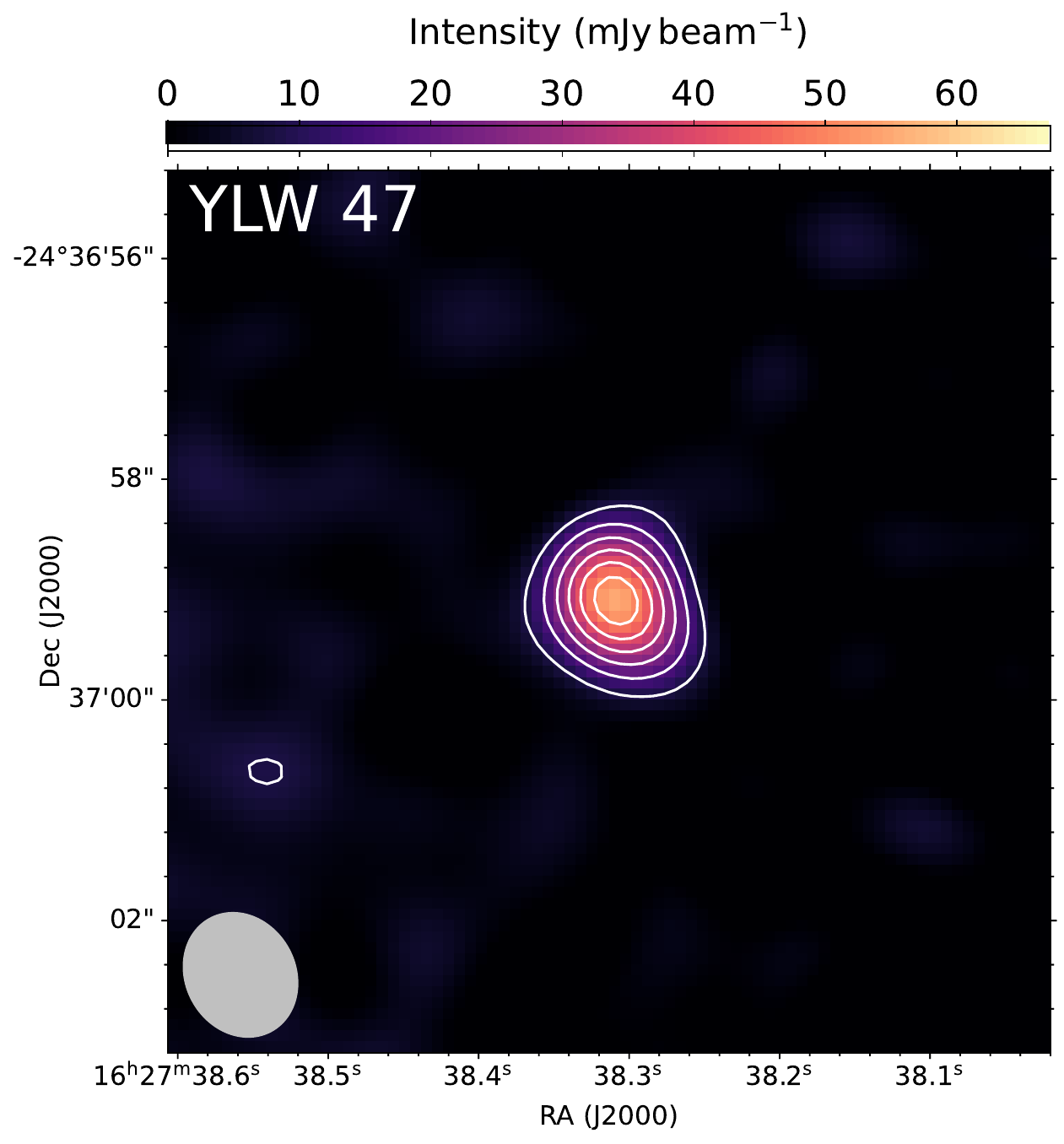}\\
  \includegraphics[width=4cm]{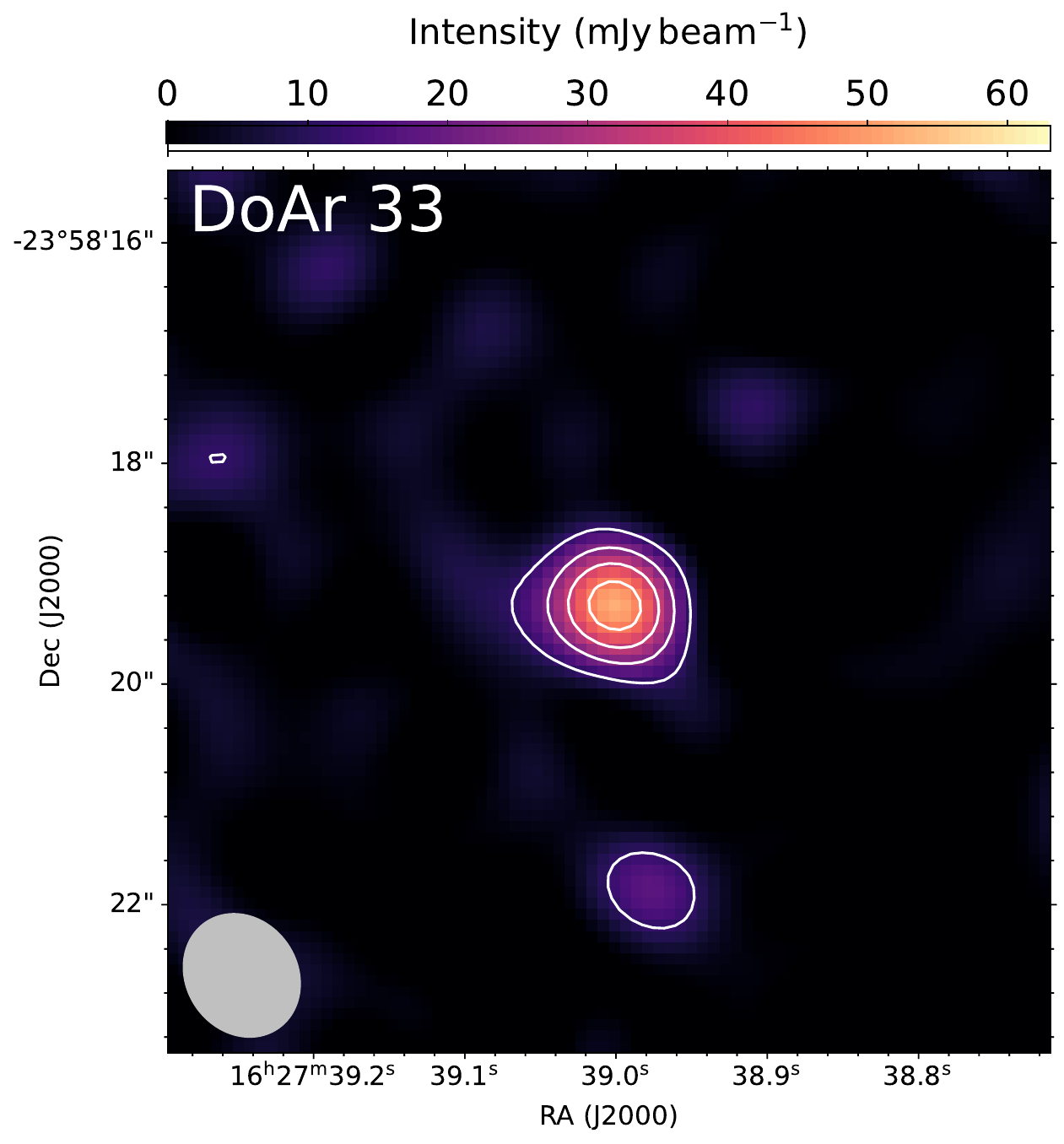} & 
  \includegraphics[width=4cm]{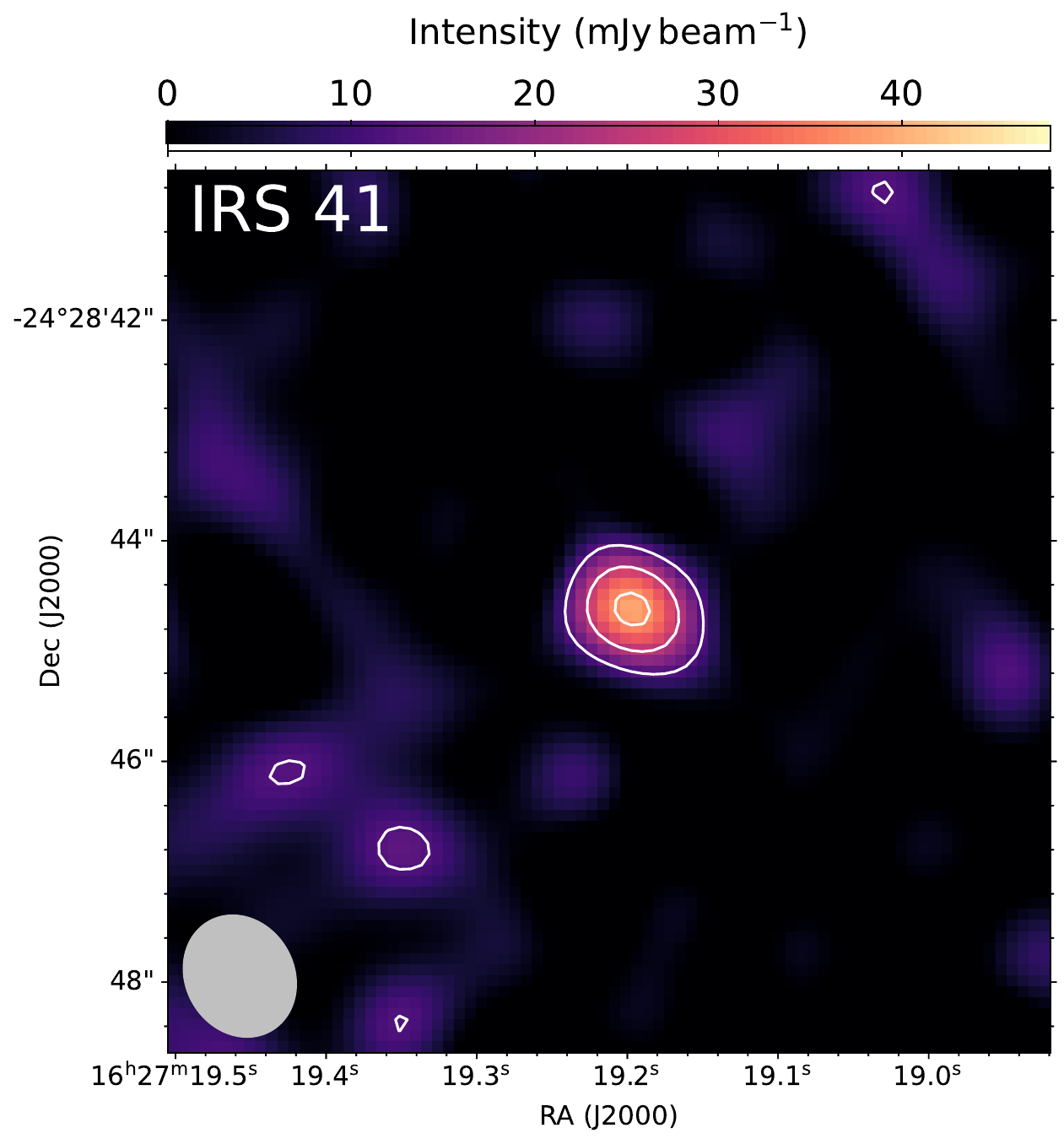} & 
  \includegraphics[width=4cm]{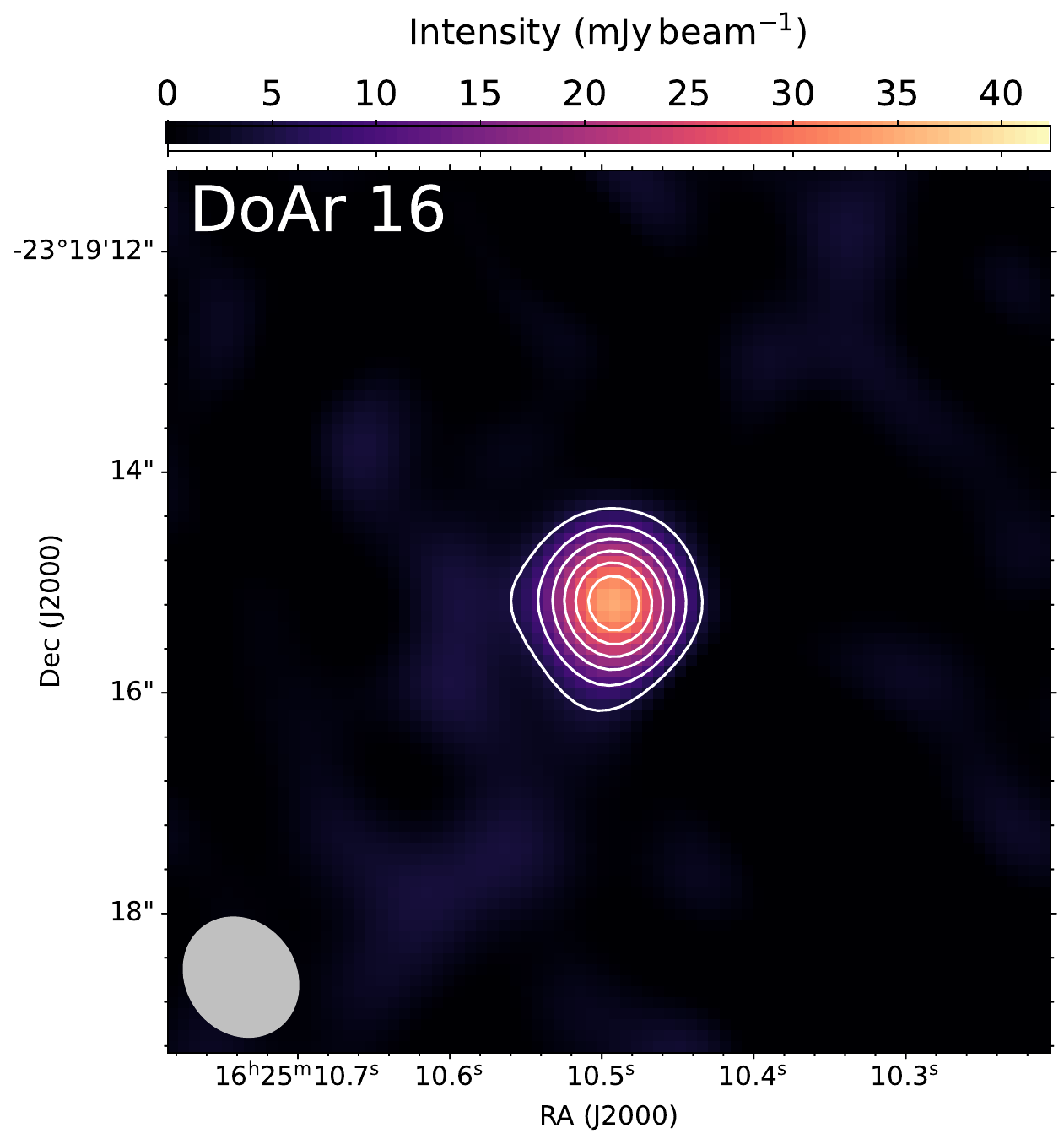}& 
  \includegraphics[width=4cm]{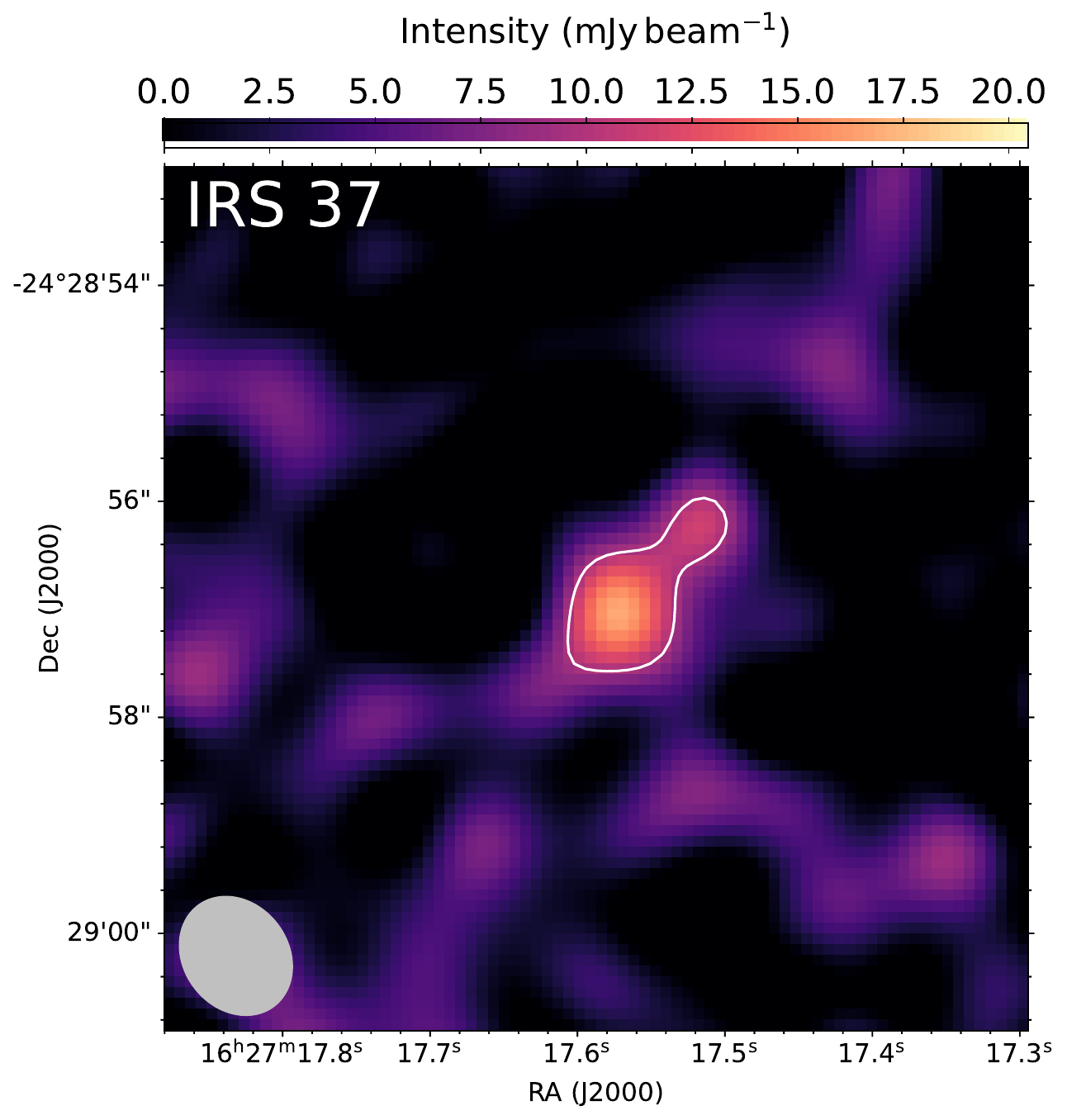}\\
  \includegraphics[width=4cm]{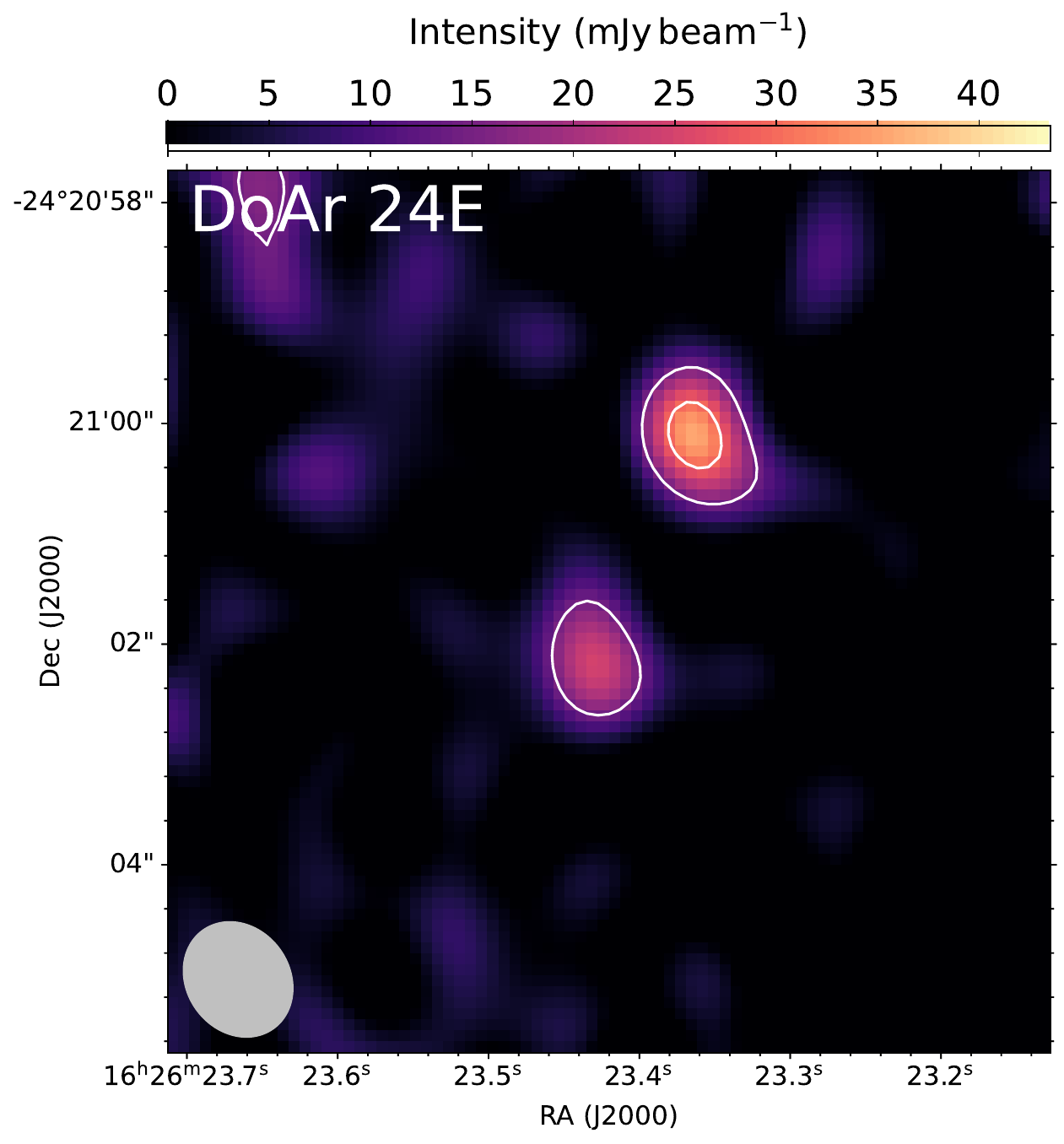} & & & \\ 
  \end{tabular}
  \caption{Images of the selected sources (Table \ref{tab:protostars}) produced from the self-calibrated data (Section \ref{sub:selcal}) at the frequency range of 286--294 GHz (Table \ref{tab:obs_summary}). Each panel shows the 8$''$-wide square region centering at the peak of the target source. The DoAr~24E panel covers the binary components DoAr~24Ea and DoAr~24Eb. Contours start at 3-$\sigma$, with 3-$\sigma$ intervals.}
  \label{fig:images}
\end{figure*}

Figure \ref{fig:images} shows the images of the selected target sources, which were produced from the gain-phase self-calibrated SMA data (Section \ref{sub:selcal}).
The images were made by \texttt{mfs} imaging (Section \ref{sub:imaging}) of the data taken in frequency range of 286--294 GHz (i.e., the lower sideband of the 302 GHz IF; Table \ref{tab:obs_summary}). 

\section{Flux Density Measurements}\label{appdx:flux}

Table \ref{tab:flux_density_summary} summarizes the results of our flux density measurements derived from visibility fittings (Section \ref{sub:flux_density}).

\section{Disk radii}\label{appdx:radii}

We quoted the $R_{\rm 95\%}$ radii of DoAr~16, YLW~8, DoAr~24Ea, and IRS~37 from \citet{Dasgupta2025ApJ...981L...4D}.
For DoAr~24Eb, based on its millimeter luminosity, we estimated the value of $R_{\rm 68\%}$ using the luminosity-radius relation reported in \citet{Hendler2020ApJ...895..126H} (their Equation 2; see also \citealt{Tripathi2017ApJ...845...44T,Andrews2018ApJ...869L..41A}).
We converted this $R_{\rm 68\%}$ to $R_{\rm 90\%}$ using Equation (6) of \citet{Hendler2020ApJ...895..126H}, which serves as an approximation of $R_{\rm 95\%}$.
For the remaining sources in Table \ref{tab:protostars}, we quoted the values of $R_{\rm 95\%}$ from \citet{Shoshi2025PASJ...77..572S}.

\begin{deluxetable*}{l cccccccc cccccccc}
\tabletypesize{\footnotesize}
\tablecolumns{9}
\tablecaption{Flux density measured at each frequency\label{tab:flux_density_summary}}
\tablehead{
\colhead{Source} &
\colhead{$F_{201\mathrm{GHz}}$} &
\colhead{$\delta_{F_{201\mathrm{GHz}}}$} &
\colhead{$F_{217\mathrm{GHz}}$} &
\colhead{$\delta_{F_{217\mathrm{GHz}}}$} &
\colhead{$F_{230\mathrm{GHz}}$} &
\colhead{$\delta_{F_{230\mathrm{GHz}}}$} &
\colhead{$F_{246\mathrm{GHz}}$} &
\colhead{$\delta_{F_{246\mathrm{GHz}}}$} &
\colhead{$F_{262\mathrm{GHz}}$} &
\colhead{$\delta_{F_{262\mathrm{GHz}}}$} &
\colhead{$F_{278\mathrm{GHz}}$} &
\colhead{$\delta_{F_{278\mathrm{GHz}}}$} &
\colhead{$F_{294\mathrm{GHz}}$} &
\colhead{$\delta_{F_{294\mathrm{GHz}}}$} &
\colhead{$F_{310\mathrm{GHz}}$} &
\colhead{$\delta_{F_{310\mathrm{GHz}}}$} \\
\colhead{} & \colhead{(mJy)} & \colhead{(mJy)} & \colhead{(mJy)} & \colhead{(mJy)} & \colhead{(mJy)} & \colhead{(mJy)} & \colhead{(mJy)} & \colhead{(mJy)} & \colhead{(mJy)} & \colhead{(mJy)} & \colhead{(mJy)} & \colhead{(mJy)} & \colhead{(mJy)} & \colhead{(mJy)} & \colhead{(mJy)} & \colhead{(mJy)}
}
\startdata
IRS 41 & 17.1 & 1.5 & 20.6 & 1.7 & 18.3 & 1.7 & 25.9 & 2.1 & 30.1 & 2.9 & 32.7 & 4.0  & 44.7 & 4.6 & 51.2 & 5.0 \\
IRS 37 & 11.2 & 1.1 & 12.3 & 1.2 & 13.2 & 1.3 & 16.3 & 1.6 & 17.2 & 2.0 & 12.8 & 2.7  & 18.7 & 2.1 & 17.6 & 2.5 \\
DoAr 16 & 15.9 & 1.1 & 21.0 & 1.5 & 19.3 & 1.5 & 22.9 & 1.6 & 27.6 & 2.3 & 29.3 & 2.5  & 37.4 & 2.4 & 35.6 & 2.5 \\
DoAr 33 & 24.9 & 2.1 & 30.5 & 2.7 & 31.2 & 2.9 & 38.9 & 3.2 & 44.8 & 4.3 & 65.7 & 7.7  & 59.4 & 4.3 & 73.8 & 6.5 \\
YLW 8 & 61.1 & 4.6 & 90.1 & 5.9 & 108.1 & 7.0 & 130.5 & 8.3 & 151.2 & 10.3 & 177.2 & 13.1  & 234.8 & 14.3 & 272.1 & 16.9 \\
GSS 39 & 210.2 & 12.0 & 256.6 & 14.6 & 273.6 & 15.7 & 339.4 & 19.3 & 403.5 & 23.2 & 417.1 & 25.2  & 518.2 & 29.8 & 558.6 & 32.6 \\
AS 206 & 49.1 & 3.1 & 55.0 & 3.4 & 68.5 & 4.2 & 75.2 & 4.6 & 86.7 & 5.8 & 96.2 & 6.8  & 107.4 & 6.7 & 129.1 & 8.3 \\
GSS 26 & 128.2 & 7.7 & 151.5 & 9.0 & 170.4 & 10.0 & 203.9 & 11.9 & 237.8 & 14.3 & 249.1 & 16.2  & 308.4 & 18.0 & 354.0 & 20.9 \\
WSB 31 & 253.0 & 14.3 & 314.0 & 17.6 & 345.4 & 19.4 & 398.5 & 22.4 & 473.2 & 26.8 & 538.3 & 31.2  & 617.0 & 34.7 & 690.3 & 39.0 \\
VSSG 1 & 68.9 & 4.5 & 91.6 & 6.1 & 89.6 & 5.8 & 114.4 & 7.5 & 120.1 & 8.2 & 153.4 & 11.7  & 195.1 & 12.3 & 207.5 & 13.7 \\
DoAr 25 & 162.1 & 9.4 & 206.2 & 11.8 & 226.6 & 13.1 & 264.1 & 15.1 & 318.3 & 18.8 & 357.9 & 22.3  & 426.2 & 24.6 & 449.9 & 26.4 \\
IRS 51 & 62.9 & 4.3 & 74.5 & 5.4 & 72.9 & 5.0 & 90.2 & 6.5 & 99.5 & 7.3 & 129.9 & 13.1  & 129.3 & 9.7 & 128.2 & 9.1 \\
YLW 47 & 27.9 & 1.9 & 32.6 & 2.1 & 37.7 & 2.5 & 45.1 & 3.0 & 43.5 & 3.0 & 43.5 & 4.1  & 57.1 & 3.6 & 63.9 & 4.2 \\
WSB 60 & 78.6 & 5.4 & 90.7 & 6.2 & 105.8 & 7.1 & 115.1 & 7.8 & 132.1 & 9.5 & 151.5 & 12.5  & 193.3 & 12.7 & 207.8 & 14.0 \\
HBC 266 & 49.3 & 4.5 & 46.4 & 3.8 & 59.9 & 4.5 & 79.8 & 6.3 & 67.3 & 5.6 & 88.0 & 9.4  & 113.5 & 9.0 & 126.0 & 9.8 \\
DoAr 44 & 61.0 & 4.5 & 73.3 & 5.1 & 86.5 & 6.1 & 102.7 & 7.1 & 109.1 & 8.3 & 142.7 & 11.8  & 158.6 & 10.8 & 170.2 & 12.4 \\
DoAr 24Ea & 17.2 & 2.2 & 19.5 & 3.1 & 23.9 & 2.9 & 16.7 & 3.1 & 28.3 & 4.4 & 37.3 & 7.0  & 36.6 & 4.9 & 37.7 & 4.6 \\
DoAr 24Eb & 17.0 & 2.2 & 12.6 & 3.0 & 23.8 & 2.7 & 16.5 & 3.4 & 22.7 & 4.2 & 27.5 & 7.2  & 24.8 & 4.9 & 37.1 & 4.5 \\
\enddata
\tablecomments{Flux density ($F_\nu$) and standard error of flux density ($\delta_{F_{\nu}}$) are represented in mJy units.}
\end{deluxetable*}


\bibliography{main}{}
\bibliographystyle{aasjournalv7}



\end{document}